\documentclass[11pt,a4paper]{revtex4}

\usepackage[british]{babel}
\usepackage[intlimits]{amsmath}
\usepackage{epsfig,amssymb,amstext,color}

\usepackage{moreverb}

\newcommand{\ket}[1]{\left\vert #1\right\rangle}
\newcommand{\bra}[1]{\left\langle #1\right\vert}
\newcommand{\braket}[2]{\left\langle #1\left
|\vphantom{{#1}{#2}}\right.#2\right\rangle}

\newcommand{\comm}[2]{\left[#1,#2\right]}

\newcommand{\abs}[1]{\left\vert #1 \right\vert}

\newcommand{\lp}{\left(}
\newcommand{\rp}{\right)}

\newcommand\avr[1]{\left\langle #1 \right\rangle}

\newcommand\x{{\mathbf r}}
\newcommand\p{{\mathbf p}}

\newcommand\DA{\Delta_\text{A}}

\newcommand\DC{\Delta_\text{C}}

\newcommand\etat{\eta_\text{t}}

\newcommand\Liou{{\mathcal L}}
\newcommand\uu{{\mathbf u}}

\newcommand\kA{k_\text{A}}

\begin{document}

\title{C++QED: An object-oriented framework for wave-function
  simulations of cavity QED systems}

\author{A. Vukics\footnote{andras.vukics@uibk.ac.at} and H. Ritsch} 

\affiliation{Institute of Theoretical Physics, University of
  Innsbruck, Technikerstrasse 25, A-6020 Innsbruck, Austria}

\begin{abstract}
We present a framework for efficiently performing Monte Carlo
wave-function simulations in cavity QED with moving particles. It
relies heavily on the object-oriented programming paradigm as realised
in C++, and is extensible and applicable for simulating open
interacting quantum dynamics in general. The user is provided with a
number of ``elements'', eg pumped moving particles, pumped lossy
cavity modes, and various interactions to compose complex interacting
systems, which contain several particles moving in electromagnetic
fields of various configurations, and perform wave-function
simulations on such systems. A number of tools are provided to
facilitate the implementation of new elements.
\end{abstract}

\maketitle

\section{Introduction}
Based on our experience gained in recent years in Monte Carlo
wave-function (MCWF) simulations of simple moving-particle cavity QED
(CQED) systems performed with low-level codes
\cite{vukics05a,vukics05b,maschler05b,vukicsthesis}, we have decided
to summarise our know-how on the problem by developing a high-level
framework for such simulations. The framework is highly modular and
therefore easy to maintain, relies solely on standard C++ programming
techniques and therefore portable, and provides an interface which is
easy to use even for those not so familiar with the theoretical models
of moving-particle CQED (\cite{domokos03} is a review of the theory
involved). Meanwhile, thanks to the optimisation mechanisms of C++
compilers, we are safe to claim not to have noticeably lost in
efficiency as compared to our previous low-level codes. Potentially,
the framework is of good use for the quantum optics community.

Simulating moving quantum particles presents many non-trivial
numerical problems especially of stability
\cite{vukics05a,vukicsthesis}. Hence, in the framework very careful
numerics is needed. Accordingly, as discussed in
App.~\ref{sec:method}, we use a slightly modified version of the
original MCWF algorithm (cf eg \cite{molmer93}) involving the use of
adaptive step-size ODE steppers and interaction picture.

At present the framework consists of three parts: The first part is
the MCWF driver (Sec.~\ref{sec:Trajectory}), which has only an
abstract view on the open system to be simulated, represented by an
abstract class. This abstract class stands at the origin of a class
hierarchy consisting the second part of the framework
(Sec.~\ref{sec:OS}). Eg a system can be an element system or a
composite system containing several element systems. The aim of the
hierarchy is to provide the user with tools to build composite systems
from several elements, and to facilitate the implementation of such
elements. Clearly, the first two parts stand quite independently of
each other and are also generally applicable. As the third part of the
framework several elements are provided at the lower levels of the
hierarchy intended as building blocks for systems of moving-particle
CQED (Sec.~\ref{sec:CQED}). The building blocks are pumped moving
particles, pumped lossy cavity modes, pumped two-level atoms, and
interactions between them eg interaction between a cavity mode and a
pumped particle moving along or orthogonal to it. This third part is
independent of the first part, but not, of course, of the second part,
the elements stemming from the same class hierarchy.

For a given system on the highest level the user is required to write
a simple driver program in C++ in which he/she defines the system to
be simulated using the elements (selecting a number of free elements
and interactions between them) and passes this system to the MCWF
driver, which then evolves the system on a number of Monte Carlo
trajectories. A description of the user interface and example drivers
are given in Sec.~\ref{sec:Drivers}.

Note that our approach here is quite different from the one presented
in \cite{schack96}. That approach is built on a hierarchy of classes
representing Hilbert space operators and state vectors, and the
application of operators on vectors is defined. Operators acting on
complex systems can then be built from elementary operators using
direct product. A similar idea is implemented in the popular Quantum
Optics Toolbox for MATLAB \cite{tan99}. Consider for a moment how this
approach could be applied for moving particles: In this case dealing
with both operators \(x\) and \(p\) cannot be avoided. A
moving-particle state vector can be stored in either representation
(the state-vector object stores in which representation it is at the
moment), and when the other operator is to be applied, an in-place
Fast Fourier Transformation (FFT) is needed. However, as our
experience shows, such a transformation always has numerical errors,
which can disturb careful statistics.

In our approach the user is provided with a much higher level
interface, our classes representing whole physical systems instead of
Hilbert space operators. This is certainly at the cost of flexibility,
but our framework does not aim at such generality as the above, since
it has been developed with a more concrete problem in mind, in
particular, CQED with moving particles. For this given problem we
consider our approach as more efficient than the above, since, as we
will show in Sec.~\ref{sec:CQED} we can completely avoid in-place FFT.

In the following we first present the highest level of the framework,
that is, the user interface, so that the reader can immediately get a
feeling about our approach. Also, by reading Sec.~\ref{sec:Drivers}
the reader can in principle already use the framework, so this can be
considered as a short write-up. This is followed by the long write-up,
the presentation of the different parts of the framework. We include
sections entitled ``Desideratum'' in which we indicate features that
would logically belong to the given part, but are as yet missing
because we have not yet needed them. These may easily be implemented
in the future.

Finally, in Sec.~\ref{sec:TestRuns}, we summarise our test runs
performed with the framework. In the Appendices we describe our
version of the MCWF method and the most important modules used in the
framework.

The source code contains more than 60 source files and a totality of
about 4000 lines, and is distributed in \texttt{tgz} format. It can be
get either from SourceForge.net at
http://sourceforge.net/projects/cppqed/ or directly from the authors.
The framework has been tested under Debian GNU/Linux and RedHat Linux
operating systems, in both cases the GNU C++ compiler has been used
for compilation.

\section{The user interface}
\label{sec:Drivers}
\subsection{Writing drivers}
\begin{table}
\begin{tabular}{l l}
\hline
\Large\texttt{Element}s \\
\large\texttt{Free}s \\
\texttt{LossyMode} & \(\DC\), \(\kappa\), \texttt{photonCutoff}\\
\texttt{PumpedLossyMode} & '', \(\eta\)\\
\texttt{MovingParticle} & \(\omega_{\text{recoil}}\), \texttt{momentumCutoff} \\
\texttt{PumpedMovingParticle} & '', \(\eta_{\text{eff}}\),
\(K_{\text{pump}}\), \texttt{pumpModeFunction} \\
\large\texttt{Interaction}s \\
\texttt{ParticleOrthogonalToCavity} & \texttt{cavity},
\texttt{pumpedParticle}, \(U_0\)\\
\texttt{ParticleAlongCavity} & \texttt{cavity}, \texttt{(pumped)particle},
\(U_0\), \(\eta_{\text{eff}}\), \(K_{\text{cavity}}\), \texttt{cavityModeFunction}\\
\texttt{ParticleCavity2D} & \texttt{cavity},
\texttt{particle}, \texttt{pumpedParticle},
\(U_0\), \(\eta_{\text{eff}}\), \(K_{\text{cavity}}\), \texttt{cavityModeFunction}\\
\texttt{ParticleTwoModes} & \texttt{particleCavity1}, \texttt{particleCavity2}\\
\texttt{IdenticalParticles} & \texttt{(pumped)particle},
\(N_{\text{particle}}\),
\texttt{vector<}\(\ket{\phi_{\text{particle}}}\)\texttt>\\
\hline
\Large\texttt{Composite} & \texttt{vector<SubsystemsInteraction>} \\
\large\texttt{SubsystemsInteraction} & 
\texttt{Interaction\&},
\texttt{vector<subsystemSequentialNumber>}\\
\hline
\Large\texttt{HS\_Vector} & \texttt{dimension}\\
\hline
\Large\texttt{Trajectory} & \(\ket{\Psi(t=0)}\), \texttt{OpenSystem\&},
\texttt{seed}, \texttt{eps}, \texttt{dplimit}\\
\hline
\end{tabular}
\caption{Classes constituting the user interface of the framework,
  with the set of elements extendable in the future at will. Next to
  each class their most important parameters are listed, these are
  explained in the text. The \texttt{Interaction}s take references to
  their subsystems as parameters --- \texttt{cavity} is an instant of
  class \texttt{(Pumped)LossyMode}, \texttt{(pumped)particle} one of
  \texttt{(Pumped)MovingParticle}, while \texttt{particleCavity} one
  of \texttt{MovingParticleCavity} cf Sec.~\ref{sec:MPC}.}
\label{tab:UI}
\end{table}

The classes a user has to know about are listed in Tab.~\ref{tab:UI}
together with the most important parameters, which will be explained
further down in the text. The set of elements for systems in
moving-particle CQED is explained in detail in Sec.~\ref{sec:CQED}.

To ease the understanding of the framework's workings example drivers
are given in Figs.~\ref{fig:1pRingCav} and \ref{fig:2p1m}. The driver
in Fig.~\ref{fig:1pRingCav} simulates one single particle moving in a
ring cavity, that is, two travelling-wave modes propagating in
opposite directions. Both modes are lossy and one of them is
pumped. In addition, the particle can also be pumped and scatter light
from the pump into the modes. The driver in Fig.~\ref{fig:2p1m}
describes two identical particles moving orthogonal to the axis of a
single-mode cavity in a standing-wave pump field.

The user has to choose an appropriate set of free systems and the
interactions between them, and instantiate the corresponding
\texttt{Free} and \texttt{Interaction} classes with the appropriate
parameters. If two elements are exactly identical, only one object is
needed. This is the case eg with several identical particles: one
instant of the \texttt{MovingParticle} class stands for all of them
(an example of this can be seen in Fig.~\ref{fig:2p1m}). 

The \texttt{Free} objects are then to be (virtually) arranged into a
sequence starting with number 0, and the user has to create a
\texttt{vector} of \texttt{SubsystemsInteraction} class objects. The
latter is a helper class for the \texttt{Composite} class, storing a
reference to an \texttt{Interaction} and the sequential number of
those \texttt{Free} objects between which the given interaction
acts. Most interactions will be between two subsystems, but we have
found cases with interactions between three or four subsystems (cf
Sec.~\ref{sec:CQED}). The \texttt{IdenticalParticles} class is an
\texttt{Interaction} between all the particles, that is, an arbitrary
number of subsystems in principle.

When giving the sequential numbers the user has to remain consistent
with the originally conceived sequence of the \texttt{Free} objects,
and the order of the subsystems in an \texttt{Interaction} object is
also important. Eg in Fig.~\ref{fig:1pRingCav} Line 19 instead of
\texttt{(pc1,1,0)} it would be an error to write \texttt{(pc1,1,2)}
because \texttt{Free} Nr.~2 is a \texttt{PumpedLossyMode} and not a
\texttt{MovingParticle}, but also \texttt{(pc1,0,1)} because
\texttt{ParticleAlongCavity} is an interaction between a
\texttt{LossyMode} and a \texttt{MovingParticle} and not vice
versa. Such errors cause an exception during the construction of the
\texttt{Composite} object.

The free systems provide helper functions to prepare state vectors (of
class \texttt{HS\_Vector}) characteristic to the given system. Eg for
a \texttt{LossyMode} object one can prepare a Fock state or a coherent
state. This, together with the possibility of making up direct
products of several state vectors, facilitates the preparation of
initial conditions. Eg in Fig.~\ref{fig:1pRingCav} Line 26 we prepare
a state in which the particle has a wave packet centred at position
\texttt{x0} with momentum \texttt{k0} and spread \texttt{xsig}, the
\(+K\) mode is in a coherent state with complex amplitude
\texttt{alpha}, and the \(-K\) with \texttt{beta}. Here again, we have
to comply with the our preconceived order of the \texttt{Free}
objects in the sequence.

As output such a program first summarises the parameters of the
system, then at certain time instants (whose frequency is specified by
the user) displays the time and the time step followed by a set of
quantum averages specified in the element system classes. At specified
time instants the whole state vector is displayed, but in practice
this can be too big to store and gain information from. An example
output is given in Fig.~\ref{fig:exampleOutput}.

\subsection{Desideratum}
With some effort the preparation of drivers could be made automatic,
such that the user is presented with a higher level interface in which
he/she specifies the system using some simple formal language, and
then the framework writes and compiles the C++ driver corresponding to
the system. A similar idea can be found implemented in the XMDS
package \cite{collecutt01}.

\section{Evolution}
\label{sec:Trajectory}
\subsection{MCWF trajectories}
\label{sec:MCWFS_Driver}
What we expect from a MCWF trajectory driver class (called
\texttt{Trajectory} in our framework); what parameters does it need
and what functionalities should it provide?

First we need to represent the state vector of the system. The most
straightforward representation is a \emph{complex packed array} (CPA),
that is, a real array, in which the real and imaginary parts of the
state-vector amplitudes are placed in alternate neighbouring
elements. In our framework the low-level notion of a CPA is furnished
with an interface class called \texttt{HS\_Vector} (for Hilbert-space
vector) supplying the operations we expect for a vector of a
Hilbert space. These include algebraic (vector-space) operations
including direct product of several vector spaces, metric operations,
and both low and high level access to amplitudes. When instantiating a
\texttt{Trajectory} the initial condition of the system has to be
given in an appropriate instant of \texttt{HS\_Vector} and this is
eventually \emph{replaced} by the driver when evolving the system.

Every system must supply an interface towards the trajectory driver
containing the operations needed to perform a MCWF step on the system
as described in App.~\ref{sec:method}. This interface is the abstract
view the driver has on the system to be simulated. In the present
framework such an abstract system is represented by an abstract class
called \texttt{OpenSystem}. The hierarchical implementation of this
interface for more and more concrete systems constitutes the main part
of the work presented here and is described in Sec.~\ref{sec:OS}. In
the C++ implementation of the object-oriented paradigm, an abstract
class cannot be instantiated but can be referred to by a reference (a
pointer), to preserve run-time polymorphism. Hence, a
\texttt{Trajectory} object takes a reference to an
\texttt{OpenSystem}.

An ODE integrator and a random-number generator are needed to perform
Step 1 and 2 of an MCWF step, respectively. These are also wrapped into
interface classes called \texttt{Evolved} and \texttt{Randomized},
respectively. At the moment, these classes are implemented using the
Gnu Scientific Library (GSL) \cite{gsl}, but here a user is free to
choose his/her own favourite library (eg Numerical Recipes) or even
hand-crafted code. To better localise object creation, only
``factory'' objects for these classes are passed to the
\texttt{Trajectory} object (for a description of the factory-class
and other programming techniques appearing in this paper see
\cite{stroustrup}).

Other important parameters are the highest allowed jump probability
\texttt{dplimit} and the relative precision for the ODE stepper
\texttt{eps}. 

The class supplies a member function called \texttt{Step} to perform
one adaptive-stepsize MCWF step on the system as follows:
\begin{enumerate}
\item Invokes the ODE stepper to evolve the state vector according to
  Eq.~(\ref{eq:DynnU}) for a suitable time interval
  \texttt{dtdid}. \(H_{\text{nH}}\) for the system is taken from the
  \texttt{OpenSystem} class.
\item Performs the additional (exact) part of the evolution as
  \(\ket{\Psi(t+\delta t)}=U^{-1}(\delta t)\ket{\Psi_{\text
  I}(t+\delta t)}\).
\item Examines whether a jump should be made. For this it uses a
  random number, \texttt{dtdid}, and a system-specific jump function
  again taken from the \texttt{OpenSystem} class.
\item The ODE stepper supplies a time step \texttt{dttry} which is
  likely to work for the \emph{next} step. The driver examines whether
  the jump probability would have overshoot \texttt{dplimit} were it
  calculated with \texttt{dttry} instead of \texttt{dtdid}. If this is
  the case, \texttt{dttry} is reduced.
\item Calculates and communicates towards the user physical
  properties of the system at the given time instant, such as the
  state vector itself and/or important quantum averages --- exactly
  what is again taken from \texttt{OpenSystem}.
\end{enumerate}

A number of helper functions are provided to take not only a step but
evolve a whole trajectory or an ensemble average of trajectories.

\subsection{Desideratum}
Other methods of wave-function simulation of open systems can be
straightforwardly added to the framework, although the
\texttt{OpenSystem} interface may need to be extended. These include
the quantum state diffusion method \cite{gisin92}, and the orthogonal
quantum jump method \cite{diosi86}. It would be advisable to keep a
common interface for the different drivers, so that the same helper
functions work for all of them.

Wave function simulations can very efficiently be done parallel. With
additional helper functions parallel execution can be easily
implemented.

\section{System hierarchy}
\label{sec:OS}
Every class derived from \texttt{OpenSystem} is an
\texttt{OpenSystem}, features the same interface, and hence can be
passed to the trajectory driver.

\begin{figure}
\centering
\includegraphics[width=16cm]{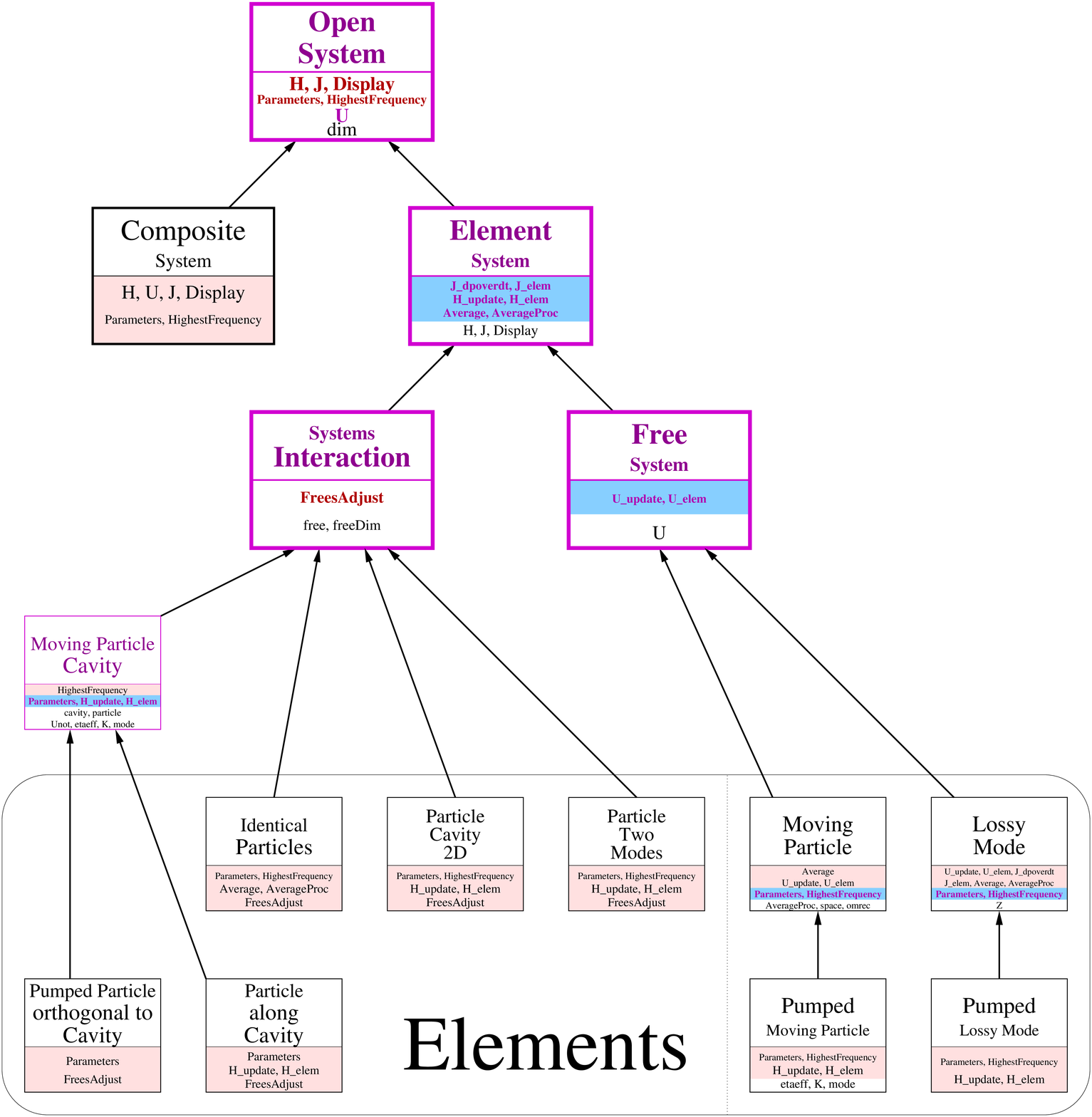}
\caption{Class inheritance hierarchy starting from the almost purely
  abstract interface \texttt{OpenSystem}, the interface that all
  simulated systems has to provide for our \texttt{Trajectory}
  driver. At the bottom of the hierarchy we have provided an example
  set of \texttt{Element}s taken from CQED with moving
  particles. These may serve as building blocks for
  \texttt{Composite}s. The colour code: magenta-framed classes are
  abstract classes, black-framed ones are concrete types; arrows
  denote class inheritance; in each class the most important functions
  are displayed --- purely virtual ones in red, virtual ones in
  magenta and concrete ones in black; the functions displayed in the
  salmon stripes belong to the private part of the class while the
  blue and white stripes refer to the protected and public part,
  respectively. The displayed functions are partly documented in the
  text, and partly in the source code.}
\label{fig:hierarchy}
\end{figure}

As indicated in Fig.~\ref{fig:hierarchy} an \texttt{OpenSystem} is
either \texttt{Composite} or \texttt{Element} system. Element systems
can be used as building blocks to compose composite systems. One may
wonder why derive also \texttt{Element} from \texttt{OpenSystem} when
elements are simple systems with known behaviour, so that one is
unlikely to wish to simulate such systems. The answer is that one
\emph{may} wish to simulate them for testing purposes when
implementing a new \texttt{Element} class. Also, this way quite an
amount of code can be reused.

An \texttt{Element}, in turn, can be either a \texttt{Free} system or
an \texttt{Interaction} of such systems. We emphasise the fact that an
\texttt{Interaction} is also an \texttt{Element}, and hence an
\texttt{OpenSystem}. One is even less likely to wish to simulate only
the interaction part of the dynamics without the free
systems: The reason for this arrangement is again code reuse.

\begin{figure}
\centering
\includegraphics[width=12cm]{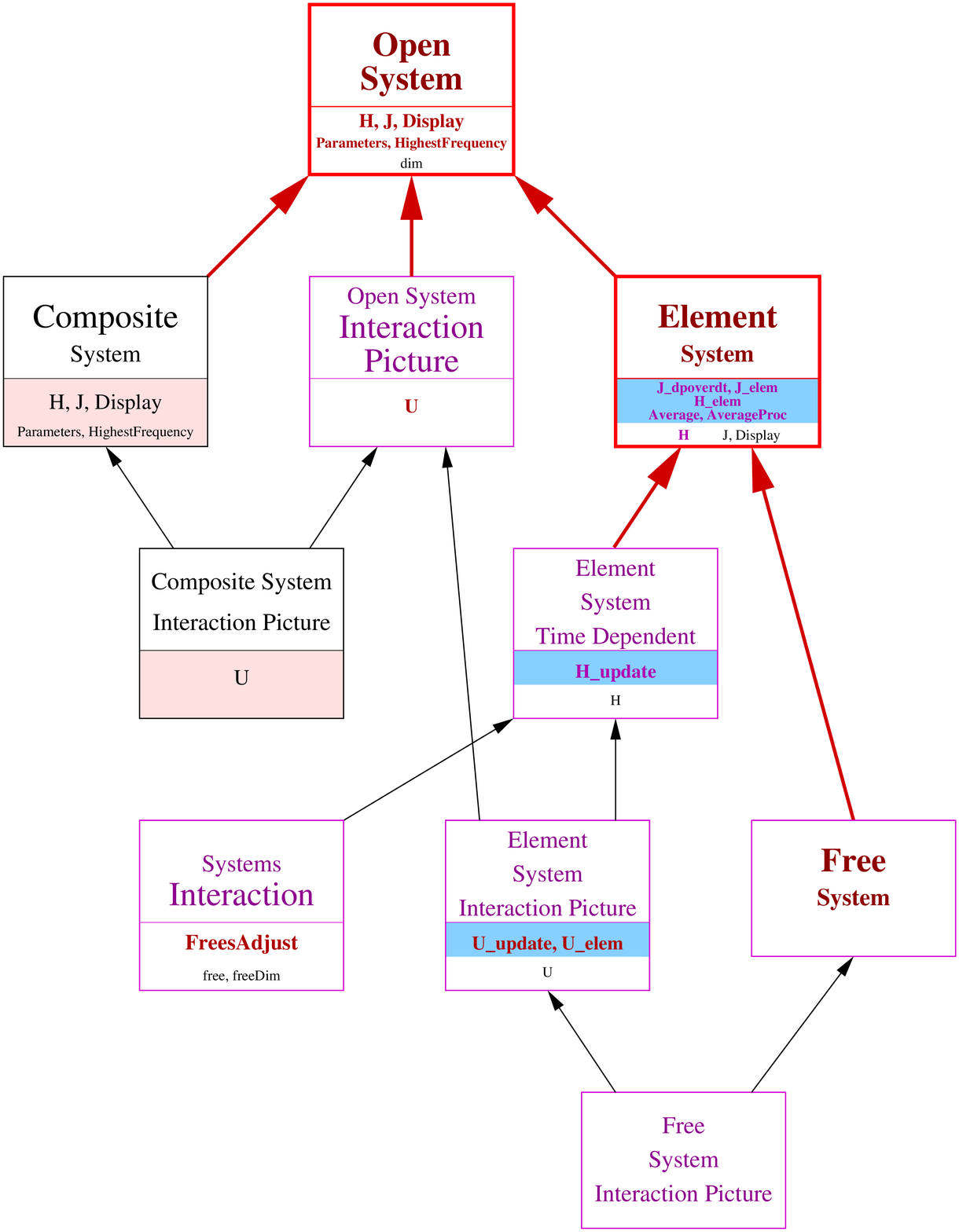}
\caption{Alternative design featuring a completely separate branch for
  systems using interaction picture. Red-framed classes are virtual
  bases, and red arrows denote virtual inheritance.}
\label{fig:hierarchy_alter}
\end{figure}

We note that we had considered the alternative design depicted in
Fig.~\ref{fig:hierarchy_alter}. Here, there is a very clear
distinction between system that use interaction picture and those that
do not. In many sense this design is more logical and attractive,
since it grasps better the structure of the problem. However, it
involves the use virtual bases, consisting a slight efficiency
overhead, and, more importantly, a bigger overhead in the complexity
of the code. We therefore eventually resorted to the first simpler
design for the testing phase.

\begin{figure}
\centering
\includegraphics[width=17cm]{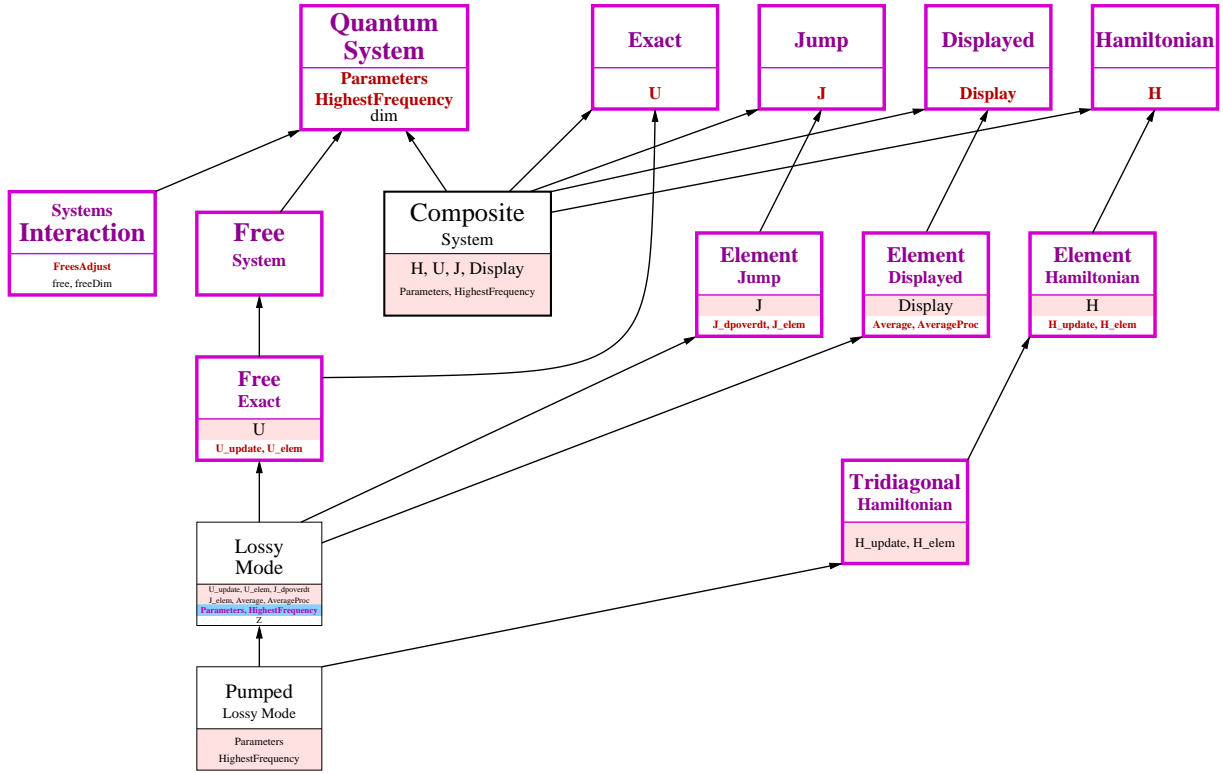}
\caption{The design actually used in the framework. The advantage over
the first design is that the fundamental functions \texttt{H},
\texttt{U}, \texttt{J}, and \texttt{Display} are declared as
\emph{pure} virtual, and therefore it is very clear which class has
implemented which function. Still, it does not use virtual bases as
the second design. The function of \texttt{Element} has ceased to
exist so this class is omitted, we have instead a set of classes
\texttt{ElementHamiltonian} etc. \texttt{Composite} then deals
separately with \texttt{Free}s and \texttt{Interaction}s. Logically,
the root of the hierarchy is not called \texttt{OpenSystem} anymore,
since the jump function is declared outside this class, but merely
\texttt{QuantumSystem}. As an example we have plotted
\texttt{LossyMode} and \texttt{PumpedLossyMode} to show how concrete
elements fit into this hierarchy. Note that eg \texttt{H} cannot even
be called for \texttt{LossyMode}, only for \texttt{PumpedLossyMode}
since the first is \emph{not} derived from the \texttt{Hamiltonian}
class.}
\label{fig:hierarchy_ultimate}
\end{figure}

The design we have found ultimately useful is, however, the one
depicted in Fig.~\ref{fig:hierarchy_ultimate}. This one unites the
advantages of the previous two, without the overhead of virtual
bases. This design is uncompromising in the sense that it is very
clearly expressed which virtual functions a class at the lower levels
of the hierarchy has to implement.

Although the underlying design in our framework is this last one, in
the following, for the sake of simplicity, and to ease the
understanding for those not so familiar with object-oriented
programming, we go on presenting the framework as if the underlying
design was the \emph{first} one. The differences are purely technical
throughout.

\subsection{\texttt{OpenSystem}}
The \texttt{OpenSystem} class is not a purely abstract one, since it
has one data member: the dimension of the system --- a parameter every
quantum system has in common. In addition it features a number of
virtual functions (function prototypes) which enable the driver class
to perform a MCWF step as described in Sec.~\ref{sec:MCWFS_Driver}. Eg
the (non-Hermitian) Hamiltonian of the system is implemented by the
function
\begin{equation*}
\texttt{void H ( double t, const double* Psi, double* dPsidt, const
CPA\_View\& V );}
\end{equation*}
The first three arguments are the expected ones: time, an array
for the state vector \(\ket\Psi\), and one for the state-vector
derivative \(d\ket\Psi/dt\). It is the last parameter that needs some
explanation. Since an \texttt{OpenSystem} can be an \texttt{Element}
system, it must be prepared to be embedded into a complex system as a
subsystem.
If so, to be able to perform the operation on the state vector of the
whole system, \texttt{H} must have some information about the
embedding complex system. As explained in App.~\ref{sec:slices}, this
information can be condensed into a set of \emph{array slice}s, which
set, in turn, is implemented by a class called \texttt{CPA\_View} in
our framework.

The other important virtual member functions are \texttt{U},
\texttt{J}, and \texttt{Display}, which take care of Phases 2, 3, and 5
of a MCWF step as described in Sec.~\ref{sec:MCWFS_Driver},
respectively. They all take arguments one would expect them to,
\emph{plus} a \texttt{CPA\_View}.

A further important virtual member function is called
\texttt{HighestFrequency}, and returns the highest characteristic
frequency in the system's time evolution --- a measure what every
dynamical system is expected to have. This is needed by the
\texttt{Trajectory} driver to initiate the ODE stepper: adaptive
step-size ODE steppers need a good guess for the initial time step to
try, which is derived by the driver from the highest characteristic
frequency of the system.

\subsection{\texttt{Element}}
\label{sec:elem}
At the level of \texttt{OpenSystem} the functions \texttt{H},
\texttt{U}, \texttt{J}, and \texttt{Display} are \emph{virtual}
functions because we can not tell what these functions are to do for a
general \texttt{OpenSystem}.

An \texttt{Element} system will be mostly embedded into a complex
system as a subsystem. As explained in detail in App.~\ref{sec:slices}
to calculate eg the Hamiltonian it has to iterate over the
state-vector slices contained by its \texttt{CPA\_View}, which
corresponds to iterate over all the possible combinations of the
quantum numbers of other subsystems --- the ``dummy'' quantum numbers
from the given subsystem's point of view, and call the same function
on the corresponding slice. Function \texttt{H} is implemented
accordingly, and class \texttt{Element} hence features the virtual
function
\begin{equation*}
\texttt{void H\_elem ( const double* Psi, double* dPsidt, const
CPA\_Slice\& S ) const;}
\end{equation*}
Note that the time argument is not passed over to
\texttt{H\_elem}. The time dependence of the original Hamiltonian
\texttt{H} is rather taken care of by another virtual function
\texttt{H\_update}, which updates the inner state of the object if it
does not correspond to the given time instant. With this method much
calculation can be saved when the same object is used to describe
several identical subsystems.

Note that \texttt{Element} is also an abstract class because although
it implements function \texttt{H} from \texttt{OpenSystem}, it
declares new virtual functions, which must be implemented further
down in the hierarchy.

\texttt{J} and \texttt{Display} are implemented along similar lines as
\texttt{H}, in both cases new virtual functions are declared. Eg
for \texttt{J} we need a function \texttt{J\_dpoverdt} which
calculates the probability of a jump per unit time in the given
subsystem, and a function \texttt{J\_elem} which actually performs the
jump on a given state-vector slice if required.

\texttt{U} is not implemented by \texttt{Element}. An \texttt{Element}
can be \texttt{Free} or \texttt{Interaction}. \texttt{U} represents
the part of the dynamics which can be exactly solved, that is, the
part of the Hamiltonian which can be diagonalised. This is possible
for some free systems, but not for interactions. Therefore \texttt{U}
is implemented only in class \texttt{Free}, along exactly the same
lines as \texttt{H} in class \texttt{Element}.

Interactions may affect the parameters of frees. A straightforward
example for this is a cavity mode whose resonance frequency is shifted
when interacting with an atom. Hence, class \texttt{Interaction}
features a virtual function called \texttt{FreesAdjust}, which
performs the required modification in the parameters of the free
systems. It is important to note that this is done at the construction
of \texttt{Composite} rather than at the construction if the given
\texttt{Interaction}. Indeed, at the construction of the interaction
we do not yet know how many times it will be applied: this becomes
clear only when we already know the layout of the whole composite
system --- in the above example the cavity frequency has to be shifted
\emph{twice} if there are two atoms instead of one.

Not every element has to implement all the virtual functions declared
in class \texttt{Element}. Eg we can easily imagine free systems whose
dynamics can be exactly solved. In this case the coherent evolution is
completely taken care of by \texttt{U}, hence \texttt{H\_elem} and
\texttt{H\_update} need not be implemented. An other common case is
when an element's dynamics is purely coherent. In this case the
functions connected to \texttt{J} are not implemented. An interesting
case is that of \texttt{IdenticalParticles} cf Sec.~\ref{sec:IdPart},
which can be considered the extreme: this class exists solely to
perform calculations in occupation-number representation, and
implements solely the functions related to \texttt{Display}.

\subsection{\texttt{Composite}}
A very important task of class \texttt{Composite} is to keep track of
its elements (frees and interactions) and their
\texttt{CPA\_View}s. The calculation of the \texttt{CPA\_View}s takes
place already at the construction of the \texttt{Composite} object.

\texttt{Composite} is a concrete type, so that it has to implement all
the virtual functions of its parent class \texttt{OpenSystem}. Eg
\texttt{H} is implemented as calling successively the \texttt{H} of
each element with the \texttt{CPA\_View} of the given element. For
this to work, it is important that the \texttt{H} functions of the
elements \emph{add} their contribution to \texttt{dPsidt} rather than
replace it. Hence with the successive calls the contributions of
elements add up, according to the model (\ref{eq:CompositeH}).

The implementation of the composite \texttt{U} and \texttt{Display} is
rather similar, only \texttt{J} needs a bit more elaboration, since
here the element \texttt{J}s should not be performed one after the
other, but a \emph{choice} has to be made as to which one (if any) to
perform. The interested reader should refer to the code to see how
this is implemented.

\subsection{Desideratum}
It is an interesting possibility, and one whose implementation should
not be too difficult in the framework to allow composite systems to be
elements of even more composite systems. This would be useful eg to
facilitate the simulation of several atoms of complex structure.

\section{Example: moving particles in cavity}
\label{sec:CQED}
\subsection{Theory}
Let us consider a single pumped two-level atom interacting with a
single pumped lossy cavity mode. Our units are chosen such that
\(\hbar=1\). Using the Jaynes-Cummings model to describe the arising
interactions, the Hamiltonian for such a system reads (\(a\) is the
cavity field operator, the \(\sigma\)s are the atomic internal
operators, \(\x\) and \(\p\) are the atomic position and momentum
operators)
\begin{subequations}
\label{eqsys:full}
\begin{equation}
\label{eq:Hamfull}
H=-\DC\,a^\dag a+i\lp \eta a^\dag-\eta^* a\rp
+\frac{\p^2}{2\mu}-\DA\,\sigma_z+
i\lp\etat^*(\x)\sigma-\etat(\x)\sigma^\dag\rp
-i \lp g(\x)\sigma^\dag a-g^*(\x) a^\dag\sigma\rp,
\end{equation}
where the terms describe free field, pumping of the mode, atomic
external and internal degrees of freedom (free and pumped), and
atom-mode interaction, respectively. The Liouvillean reads
\begin{equation}
\label{eq:Lioufull}
\Liou\rho=\kappa\lp2a\rho a^\dag-\comm{a^\dag a}{\rho}_+\rp+
\gamma\lp2\int d^2\uu\,N(\uu)\,
\sigma e^{-i\kA\uu\x}\rho\,e^{i\kA\uu\x}\sigma^\dag-
\comm{\sigma^\dag \sigma}{\rho}_+\rp,
\end{equation}
\end{subequations}
where the first term describes cavity decay and the second one atomic
spontaneous emission. 
The second term contains momentum recoil due to spontaneous
emissions. The unit vector \(\uu\) is the direction of the
spontaneously emitted photon, and \(N(\uu)\) the direction
distribution characteristic to the given atomic transition.

The operator (\ref{eq:Lioufull}) conforming with Eq.~(\ref{eq:mm}) we
can immediately read the necessary jump operators for this
system. There is one for cavity decay and an infinite set parametrised
by \(\uu\) for atomic decay:
\begin{equation}
J_\text{C}=\sqrt{2\kappa}\,a,\quad 
J_\text{A}(\uu)=\sqrt{2\gamma}\,e^{-iK\uu\x}\,\sigma.
\end{equation}
We introduce \(Z_{\text C}=\kappa-i\DC\), \(Z_{\text A}=\gamma-i\DA\).
The non-Hermitian Hamiltonian is obtained by replacing \(\DC\) with
\(iZ_{\text C}\) and \(\DA\) with \(iZ_{\text A}\) in
Eq.~(\ref{eq:Hamfull}). 

In the limit of large atomic detuning \(\DA\) the atomic internal
degree of freedom \(\sigma\) can be adiabatically eliminated, as
described in Refs.~\cite{domokos03,vukics05b}:
\begin{equation}
\label{eq:adiab}
\sigma\approx\frac{g(\x)\,a+\etat(\x)}{i\DA-\gamma}.
\end{equation}
In this limit the atomic spontaneous emission can be neglected in most
cases of interest. We will resort to this approximation to simplify
the discussion. Putting \(\gamma=0\) leaves us with only one jump
operator 
\begin{subequations}
\label{eqsys:effMCWF}
\begin{equation}
J_\text{C}=\sqrt{2\kappa}\,a.
\end{equation}
We plug (\ref{eq:adiab}) into (\ref{eq:Hamfull}). We take
\(g(\x)=gf(\x)\) and \(\etat(\x)=\etat\zeta(\x)\), and assume that
\(g\) and \(\etat\) are real (the possibility of their being complex
is investigated in \cite{vukics04}). We obtain the following effective
non-Hermitian Hamiltonian
\begin{multline}
\label{eq:Hameff}
H_\text{eff}=
-\lp i Z_{\text{C}}-U_0\abs{f(\x)}^2\rp a^\dag a+
i\lp\eta a^\dag-\eta^* a\rp
+\frac{\p^2}{2\mu}+\eta_\text{eff}\abs{\zeta(\x)}^2\\+
\text{sign}(U_0)\sqrt{U_0\,\eta_\text{eff}}
\lp 
f^*(\x)\,\zeta(\x)\,a^\dag+\text{h.c.}\rp,
\end{multline}
\end{subequations}
with \(U_0=\abs g^2/\DA\), \(\eta_{\text{eff}}=\abs\etat^2/\DA\).

The following set of elements realizes the system
(\ref{eqsys:effMCWF}). An important restriction whose reason will
become apparent later in this section is that the mode functions are
restricted to one dimension and either standing- or travelling-wave
modes:
\begin{equation}
\label{eq:ModeFunctionType}
f(\x),\,\zeta(\x)=m(\xi)\equiv
\left\{
\begin{matrix}
\sin(K\xi)\\
\cos(K\xi)\\
e^{\pm iK\xi}
\end{matrix}
\right.,\quad
\xi=x,y,z.
\end{equation}

\subsection{Free elements}
These classes implement \texttt{H\_elem}, \texttt{H\_update},
\texttt{J\_dpoverdt}, \texttt{J\_elem}, and the functions connected to
\texttt{Display}: \texttt{Average} and \texttt{AverageProc} from
parent class \texttt{Element}, and \texttt{U\_elem}, \texttt{U\_update}
from parent class \texttt{Free}. Their functionality is summarised in 
Tab.~\ref{tab:CQEDfrees}.

\begin{table}
\centering
\begin{tabular}{l | l | l 
    | l l | l}
\texttt{Free} & 
\(U(t)\) & 
\(H\) & 
\texttt{J\_dpoverdt} &
\texttt{J\_elem} & Display\\
\hline
\hline
\texttt{LossyMode} & 
\(\exp\lp-Z_\text{C}t\,N\rp\)&
\(\varnothing\) & 
\(2\kappa\,N\) & a &
\(\avr{N}\), \((\Delta N)^2\), \(\avr{a}\)\\
\texttt{PumpedLossyMode} & 
\(\Uparrow\) &
\(i\lp\eta a^\dag-\eta^*a\rp\) & 
\(\Uparrow\) &
\(\Uparrow\) &
\(\Uparrow\)\\
\hline
\texttt{MovingParticle} & 
\(\exp\lp-i\omega_\text{rec}t\,k^2\rp\) & 
\(\varnothing\) & \(\varnothing\) & \(\varnothing\) & 
\(\avr{k}\), \((\Delta k)^2\), \(\avr{x}\), \(\Delta x\)\\
\texttt{PumpedMovingParticle}& 
\(\Uparrow\) &
\(\eta_\text{eff}\abs{m(\xi)}^2\) & 
\(\Uparrow\) &
\(\Uparrow\) &
\(\Uparrow\)\\
\end{tabular}
\caption{Summary of the free elements' functionality, fully exposed in
  the text. \(\Uparrow\) indicates that the given function is
  inherited from the parent class. \(N=a^\dag a\) is the photon number
  of the mode.}
\label{tab:CQEDfrees}
\end{table}

\subsubsection{\texttt{(Pumped)LossyMode}}

Class \texttt{LossyMode} implements the dynamics of a free lossy
(cavity) mode. Its parameters are the detuning between the driving and
the cavity resonance \(\DC\), the cavity decay rate \(\kappa\), and the
photon number cutoff. 

The non-Hermitian Hamiltonian can be diagonalised
exactly, so that \texttt{H\_elem} needs not be implemented while
\texttt{U\_elem} is implemented as applying
\(U(t)=\exp\lp-Z_\text{C}t\,a^\dag a\rp\) on the state vector slice.

A \texttt{PumpedLossyMode} has the additional parameter \(\eta\).
Here only \texttt{H\_update} and \texttt{H\_elem} needs to be
implemented to apply the Hamiltonian 
\begin{equation}
H_{\text I}(t)=i U^{-1}(t)\lp\eta a^\dag-\eta^* a\rp U(t)=
i\lp\eta e^{Z_\text{C}t}a^\dag-\eta^*e^{-Z_\text{C}t}a\rp. 
\end{equation}
Since pumping does not
affect the remaining part of the dynamics, all the other functions are
exactly the same as for \texttt{LossyMode}, and
\texttt{PumpedLossyMode} indeed has access to these functions:
``inherits'' them from the parent class \texttt{LossyMode}. This is
the reason why in the class inheritance hierarchy in
Fig.~\ref{fig:hierarchy} \texttt{PumpedLossyMode} stems from
\texttt{LossyMode}. Clearly, this technique can be applied to reuse a
lot of code, and has indeed been applied throughout in our framework.

\subsubsection{\texttt{(Pumped)MovingParticle}}
A similar relationship exists between \texttt{MovingParticle} and
\texttt{PumpedMovingParticle}. \texttt{MovingParticle} implements the
dynamics of a free quantum mechanical particle moving in 1D, with
Hamiltonian \(H=p^2/(2\mu)\). This Hamiltonian is most conveniently
implemented in momentum basis. For the numerics the momentum basis
must be discrete, which amounts to some finite quantisation
volume (length). Our choice of units is such that the smallest momentum is
\(\Delta k=1\), that is, the quantisation length is \(2\pi\). It is
easy to see that the use of discrete momentum basis entails periodic
boundary condition at the borders of the quantisation length. The
parameters are the recoil frequency
\(\omega_\text{rec}\equiv\hbar\,\Delta k^2/(2\mu)=1/(2\mu)\) and the
spatial resolution. The latter has to be an integer power of 2 to be
able to perform radix-2 FFT on the state vector. With our units
\(H=\omega_\text{rec} k^2\) with operator \(k\equiv p/(\hbar\,\Delta k)=p\).

The Hamiltonian is diagonal in the momentum basis, and is quadratic in
the momentum. According to our experience, the second property makes it
essential to use interaction picture because the quadratic growth of
the frequency is too quick for the stepper routine and results in
instabilities for the higher momentum components.

According to this discussion class \texttt{MovingParticle} implements
\(U(t)=\exp\lp-i\omega_\text{rec}t\,k^2\rp\), which is diagonal in
momentum basis. The quantum averages calculated and communicated
towards the user are: \(\avr k\), \(\avr{k^2}-\avr k^2\) (proportional
to the kinetic temperature of the particle), \(\avr x\),
\(\sqrt{\avr{x^2}-\avr x^2}\). This means that at each call of
\texttt{Display} for the class, a Fourier transformation has to be
performed on a copy of the state vector to calculate the averages of
operator \(x\). This is done using the radix-2 FFT routine supplied by
GSL, but here again the user is free to use his/her own favourite
routine. We emphasise, however, that the time evolution is performed
purely in momentum representation, nor is our \texttt{Trajectory}
driver prepared to perform FFT during evolution. When FFT is performed
at all, it is on a \emph{copy} of the state vector, not an in-place
transformation. Hence, we avoid numerical errors accumulating in the
state vector, and also save the inverse transformation (although we
lose time by copying).

\texttt{PumpedMovingParticle} implements the Hamiltonian \(H_{\text
I}(t)=\eta_\text{eff}\,U^{-1}(t)\abs{\zeta(\x)}^2 U(t)\). This has to
be done in momentum space as well, therefore it pays to choose
\(\zeta(\x)\) such that it be easy to calculate its action on the
state vector in momentum space. This brings us back to the restriction
(\ref{eq:ModeFunctionType}): the action of \(e^{iK\xi}\) is very easy
to calculate as it simply amounts to a shift by \(K\) in momentum
space. For \(\zeta(\x)=e^{\pm iK\xi}\) the Hamiltonian is constant,
while for \(\zeta(\x)=\sin(K\xi),\,\cos(K\xi)\), it is proportional to
\(\mp\cos(2K\xi)/2\), respectively, after dropping the constant
term. This gives
\begin{equation}
H_{\text I}(t)=
\mp\frac{\eta_\text{eff}}2\,U^{-1}(t)\,\cos(2K\xi)\,U(t)=
\mp\frac{\eta_\text{eff}}4\,
\lp
e^{-4K\omega_\text{rec}(K-k)}e^{2iK\xi}+
e^{-4K\omega_\text{rec}(K+k)}e^{-2iK\xi}
\rp.
\end{equation}
It becomes clear how huge we gain by using interaction picture in this
case. The Hamiltonian is time dependent now, but the oscillation
frequency grows only linearly with \(k\) instead of the quadratic
growth mentioned above.

\subsection{Interaction elements}
The functionality of these classes is summarised in Tab.~\ref{tab:CQEDinteractions}.

\begin{table}
\centering
\begin{tabular}{l | l 
    | l}
\texttt{Interaction} & 
\(H\) & 
Display\\
\hline
\hline
\texttt{ParticleOrthogonalToCavity} & 
\(A\lp m(\xi)\,a^\dag+\text{h.c.}\rp\)
& \(\varnothing\)\\
\texttt{ParticleAlongCavity} & \(U_0\abs{m(\xi)}^2 N\) + '' & \(\varnothing\)\\
\hline
\texttt{ParticleCavity2D} & 
\(U_0\abs{m_1(\xi_1)}^2 N+A\lp
m_1^*(\xi_1)\,m_2(\xi_2)\,a^\dag+\text{h.c.}\rp\) & \(\varnothing\) \\
\texttt{ParticleTwoModes} & \(\sqrt{U_{01}U_{02}}
\lp m_1^*(\xi_1)\,m_2(\xi_2)\,a_1^\dag\,a_2+\text{h.c.}\rp\) & \(\varnothing\)\\
\hline
\texttt{IdenticalParticles} & \(\varnothing\) & \(\braket{2,0}\Psi\),
\(\braket{1,1}\Psi\), \(\braket{0,2}\Psi\) \\
\end{tabular}
\caption{Summary of the interaction elements'
functionality. \(N=a^\dag a\) is again the photon number,
\(A=\text{sign}(U_0)\sqrt{U_0\,\eta_\text{eff}}\).}
\label{tab:CQEDinteractions}
\end{table}

\subsubsection{\texttt{Particle(Orthogonal/Along)Cavity}}
\label{sec:MPC}
These classes implement the interaction Hamiltonians between a cavity
mode and a particle moving in 1D, either in a direction orthogonal to the
cavity axis, or the direction along it, respectively. 

Hence, \texttt{ParticleOrthogonalToCavity} implements
\begin{subequations}
\begin{equation}
H=\text{sign}(U_0)\sqrt{U_0\,\eta_\text{eff}}
\lp\zeta(\xi)\,a^\dag+\zeta^*(\xi)\,a\rp,
\end{equation}
which describes atomic stimulated absorption of a photon from the
atomic pump and stimulated reemission into the cavity mode or vice
versa. \texttt{ParticleAlongCavity} implements
\begin{equation}
\label{eq:HamPAC}
H=U_0\abs{f(\xi)}^2 a^\dag a+
\text{sign}(U_0)\sqrt{U_0\,\eta_\text{eff}}
\lp f^*(\xi)\,a^\dag+f(\xi)\,a\rp,
\end{equation}
\end{subequations}
where the first term describes atomic stimulated absorption from the
cavity mode followed by stimulated reemission into the same mode. 

These Hamiltonians are also implemented in interaction picture. Note
that the first Hamiltonian is formally identical to the second term of
the second Hamiltonian. Therefore it pays to implement this term
already in a higher level in the hierarchy, so that both of these
classes have access to it. This is done by the class
\texttt{MovingParticleCavity} which, as we see in
Fig.~\ref{fig:hierarchy} is a parent class of both.

\texttt{ParticleAlongCavity} is either instantiated with a
\texttt{MovingParticle} and an explicitly supplied parameter
\texttt{etaeff}, or with a \texttt{PumpedMovingParticle}, in which
case the \texttt{etaeff} parameter is taken from this latter
class. The first case describes the situation when the particle pump
is aligned orthogonally to the cavity axis, while the second case when
it is along the axis, so that the particle, which is also moving along
the axis, feels the pump potential as well.

The virtual function \texttt{FreesAdjust} defined in
\texttt{Interaction} is implemented so that the cavity frequency is
shifted by the interaction with the particle. In the orthogonal case
this is fairly straightforward: the shift \(\DC\to\DC-U_0\) is
applied. In fact, the user has the choice whether it should be applied
or not, in the latter case \(\DC\) stands for the \emph{shifted}
frequency. With \texttt{ParticleAlongCavity}, the situation is
somewhat more involved because the shift depends on the cavity mode
function: for \(f(\xi)=e^{\pm iK\xi}\) the shift has to be done by
\(U_0\), while in the \(f(\xi)=\sin(K\xi),\,\cos(K\xi)\) case by
\(U_0/2\) since in this case the first term of the Hamiltonian
(\ref{eq:HamPAC}) reads \(U_0/2\lp1\mp\cos(2K\xi)\rp a^\dag a\).

\subsubsection{\texttt{ParticleCavity2D}}
This class is an \texttt{Interaction} between three subsystems, and
implements the Hamiltonian
\begin{equation}
H=U_0\abs{f(\xi_1)}^2 a^\dag a+
\text{sign}(U_0)\sqrt{U_0\,\eta_\text{eff}}
\lp f^*(\xi_1)\,\zeta(\xi_2)\,a^\dag+f(\xi_1)\,\zeta^*(\xi_2)\,a\rp,
\end{equation}
which describes the situation when the pumped particle is moving in
two dimensions. One of the dimensions is taken care of by a
\texttt{MovingParticle} class and the other one by a
\texttt{PumpedMovingParticle} class --- as mentioned above these
classes implement \emph{one single} spatial degree of freedom each.

\subsubsection{\texttt{ParticleTwoModes}}
It is easy to see that if we have several cavity modes then instead of
(\ref{eq:adiab}) we have
\begin{equation}
\sigma\propto\sum_i g_i(\x)\,a_i+\etat(\x).
\end{equation}
In the effective Hamiltonian (\ref{eq:Hameff}) this creates terms like
\begin{equation}
H\propto f^*(\xi_1)\,f(\xi_2)\,a_1^\dag a_2 + \text{h.c.},
\end{equation}
which describes atomic stimulated absorption of a photon from one mode
and stimulated reemission into the other mode.

This cannot be described with the classes we have so far, so we need
one more class \texttt{ParticleTwoModes} to cover this case as
well. This closes our set of classes needed to build composite systems
of an arbitrary number of (pumped) moving particles and (pumped) lossy
cavity modes of different spatial configurations complying with the
model (\ref{eqsys:effMCWF}).

\texttt{ParticleTwoModes} is an interaction between \emph{four}
subsystems, but the two spatial degrees of freedom can be the
same. This describes the case of a linear cavity sustaining two modes
and one particle moving along it.

\subsubsection{\texttt{IdenticalParticles}}
\label{sec:IdPart}
An interesting feature of our framework is that if we have several
identical particles, it is very easy to switch between their being
considered as bosons or fermions, or even distinguishable
particles. All we have to do is to prepare the initial condition with
the appropriate symmetry with respect to the swapping of two
particles. This symmetry is then conserved during evolution.

If we consider our particles as indistinguishable, we might want to
perform calculations in some occupation-number basis. This is
facilitated by the \texttt{IdenticalParticles} class, which is an
\texttt{Interaction} between several identical particles, which are
therefore described by one single object of class
\texttt{(Pumped)MovingParticle}. At its construction, an
\texttt{IdenticalParticles} takes a reference to such a particle
object, the number of particles, and a set of single-particle state
vectors. It then constructs the occupation number basis and
\texttt{Display} is implemented such that the complex amplitudes in
this basis are calculated and communicated towards the user. Of course
this makes sense only if the single-particle state vectors are
pairwise orthogonal.

Eg for two particles and two state vectors \(\ket{\phi_1}\) and
\(\ket{\phi_2}\) the occupation-number basis for bosons looks like
\begin{subequations}
\begin{align}
\ket{2,0}&=\ket{\phi_1}\otimes\ket{\phi_1},\\
\ket{1,1}&=\frac1{\sqrt{2}}\lp\ket{\phi_1}\otimes\ket{\phi_2}+\ket{\phi_2}\otimes\ket{\phi_1}\rp,\\
\ket{0,2}&=\ket{\phi_2}\otimes\ket{\phi_2},
\end{align}
\end{subequations}
and \texttt{Display} then displays the complex amplitudes
\(\braket{2,0}\Psi\), \(\braket{1,1}\Psi\), and \(\braket{0,2}\Psi\).

In the case of indistinguishable particles it makes no sense to
calculate the quantum averages for each of them separately because due
to the symmetry all will be equal. Therefore,
\texttt{IdenticalParticles} implements \texttt{FreesAdjust} such that
the \texttt{Display} of the particles is switched off, and taken over
by \texttt{IdenticalParticles}.

\subsection{Desideratum}
We note that the above description of \texttt{IdenticalParticles}
reflects the ``ideal state'' of the class, which allows it to be used
completely generally. Clearly, for several particles and
single-particle states the implementation of this involves an amount
of combinatorics, and has not yet been done. Instead, in the first
release of the framework \texttt{IdenticalParticles} is an interaction
between \emph{two} atoms, and calculates \(\avr{n_1n_2}\), where
\(n_1\) is the number of particles at \(x<0\) and \(n_2\) at
\(x>0\). Why this is useful in some cases is explained in
\cite{maschler05b}.  Of course, this restriction of
\texttt{IdenticalParticles} does not mean that the framework can not
be used to simulate as many particles as wanted.

Atomic spontaneous emission is not implemented. In the above discussed
model, where the atomic internal dynamics is eliminated, the
implementation of this is rather involved, eg the jump operators have
to be implemented by the interaction classes since they contain both
operators \(x\) and \(a\) \cite{vukics05a}. A physical problem with
the spontaneous emission is that in the far detuned regime its rate is
given by \(\Gamma_0=\gamma\,g^2/\DA^2\), which is much smaller than
the other frequencies of the system. It therefore adds a new, very
slow relaxation time scale to the system, which makes the simulations
very long, practically unmanageable. 

It is interesting to note that when implementing spontaneous emission,
class \texttt{IdenticalParticles} gains physical significance: it has
to ensure that the particle jump operators do not modify the state
vector's symmetry with respect to particle exchange.

The next step in the development will be the addition of the two-level
atom to the framework. This entails a number of new interaction
elements, eg the term
\(i\lp\etat^*(\x)\sigma-\etat(\x)\sigma^\dag\rp\) of Hamiltonian
(\ref{eq:Hamfull}) will be an \texttt{Interaction} between a two-level
atom and one or several spatial degrees of freedom
(\texttt{MovingParticle}s).

\section{Test runs}
\label{sec:TestRuns}
Testing is difficult in our case because the behaviour of the system
we aim to simulate, that is, the coupled open quantum dynamics of
several particles and lossy cavity-field modes is largely unknown, and
constitutes an extremely rich area of active physical research --- the
framework is intended as a tool for this research.

Of course, utilities like \texttt{HS\_Vector}, \texttt{Evoled},
\texttt{Randomized}, and maybe even \texttt{Trajectory} can be tested
separately. Free elements should not present too much problem
either. Interactions are, however, more problematic.

Our principle for testing interaction elements was to find parameter
regimes where the action of one subsystem on the other(s) is very
strong, but the back-action is negligible.

\begin{figure}
\centering
\begin{tabular}{l  r}
(a) & (b)\\
\includegraphics[width=5.5cm,angle=-90]{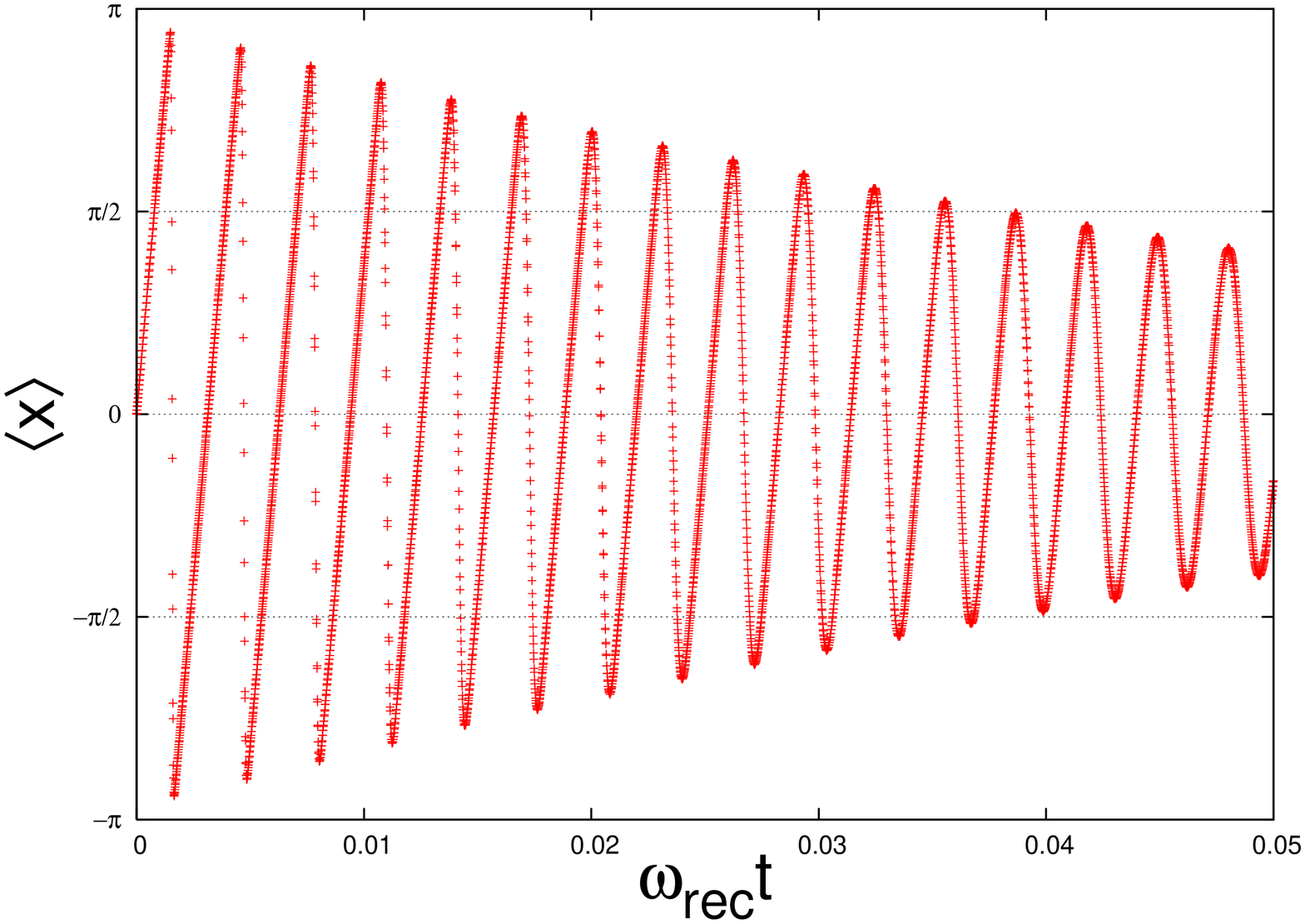} &
\includegraphics[width=5.5cm,angle=-90]{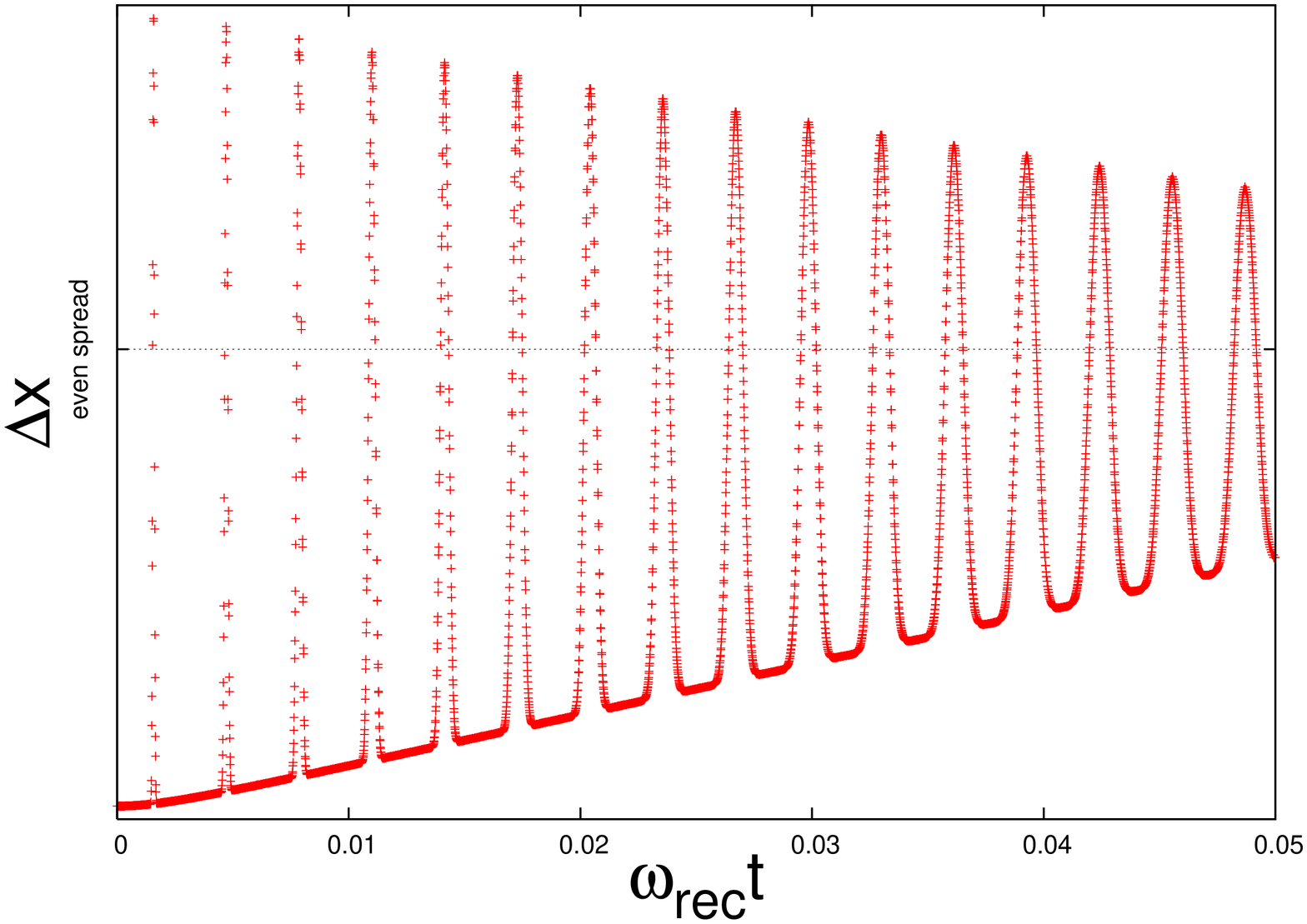}\\
(c) & (d)\\
\includegraphics[width=5.5cm,angle=-90]{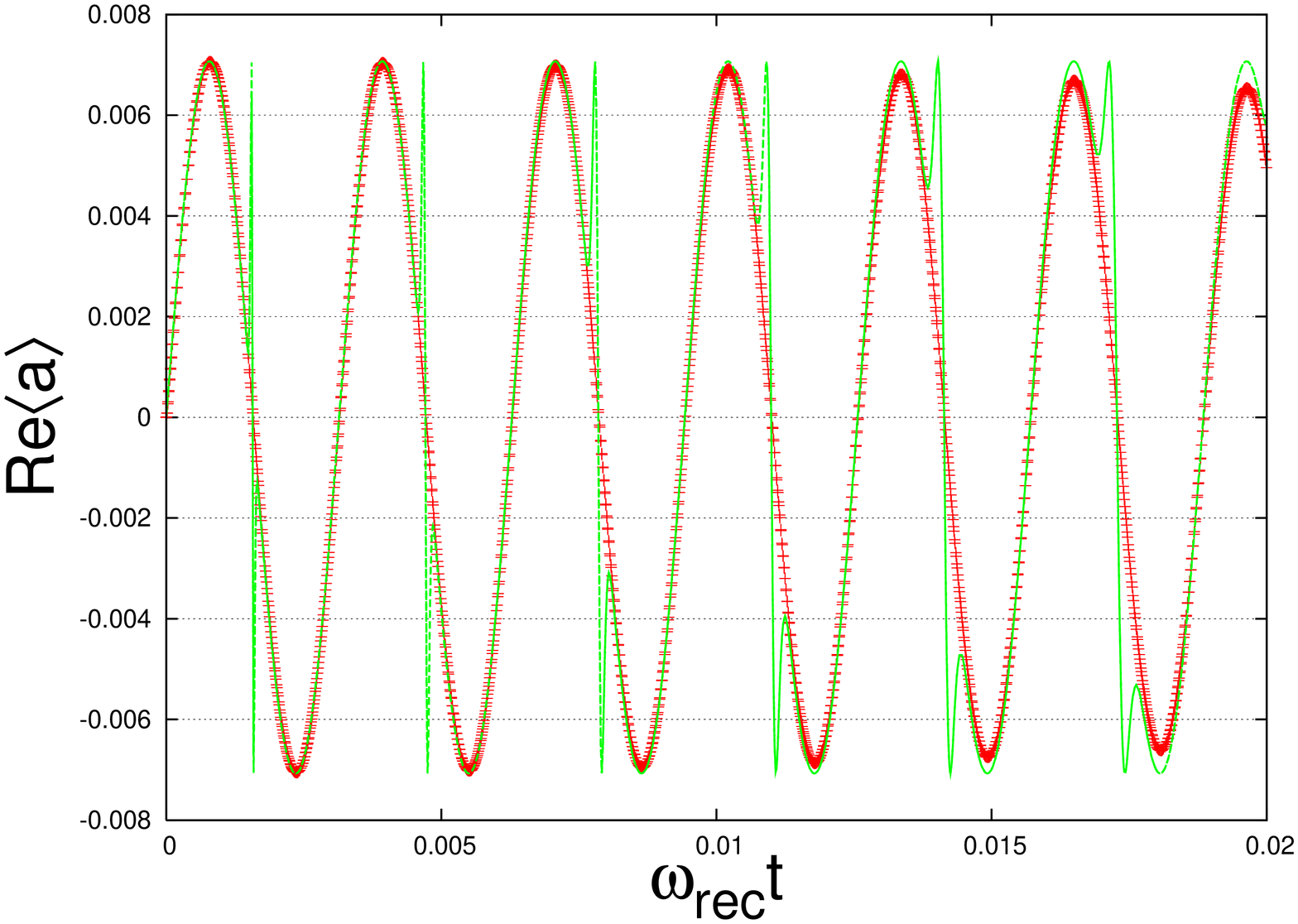} & \includegraphics[width=5.5cm,angle=-90]{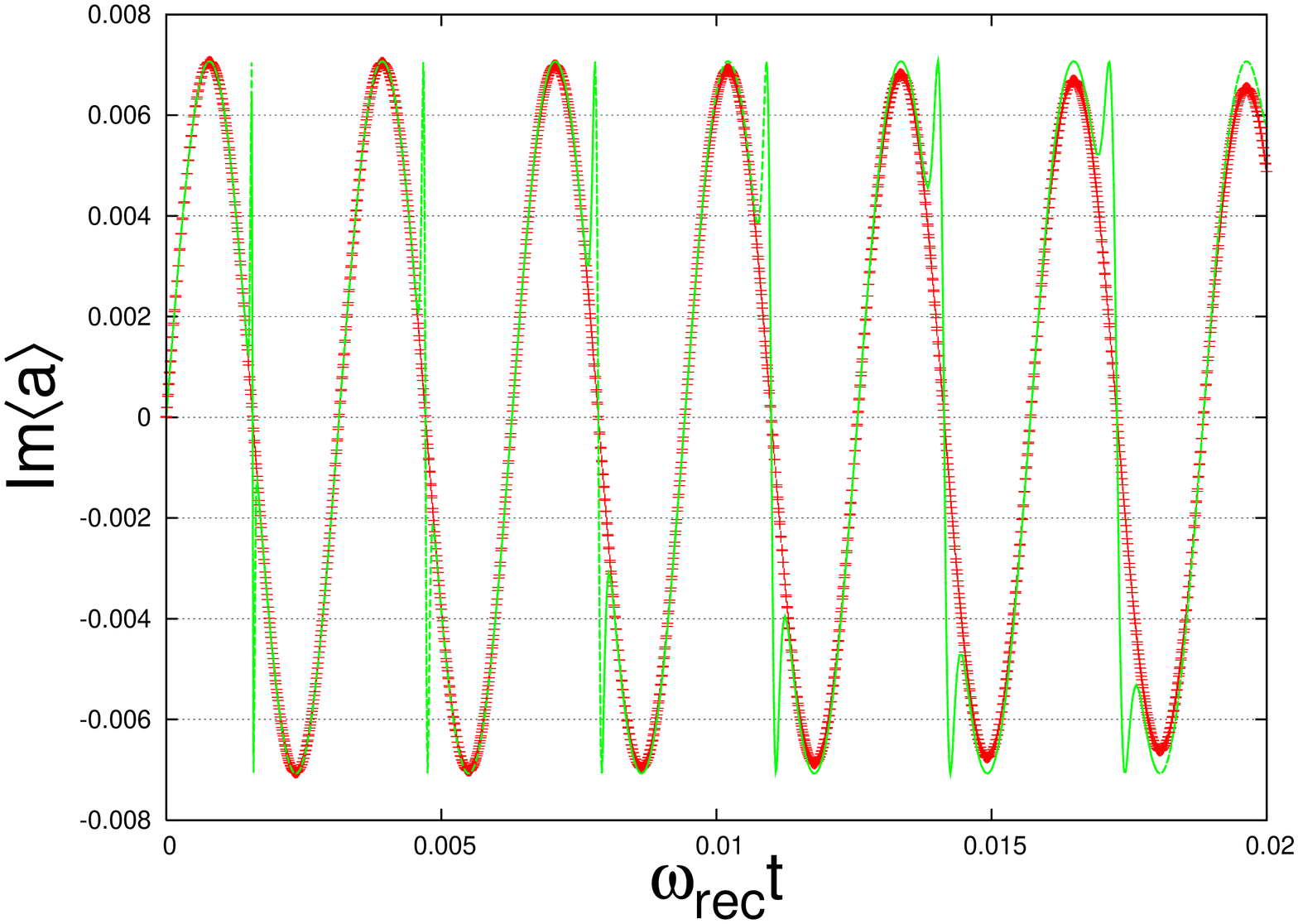}\\
\end{tabular}
\caption{Massive pumped particle moving quickly in a direction
  orthogonal to the axis of a cavity. (a) Expectation value of the
  atom's position. Each time the atom goes out of the quantisation
  volume at \(x=\pi\), it comes back in at \(x=-\pi\) due to periodic
  boundary condition. (b) Spread of the atomic wave packet. (c) \& (d)
  Real and imaginary part of the scattered field in the cavity, the
  green lines corresponding to the estimation
  (\ref{eq:ClassicalField}).}
\label{fig:TestRun}
\end{figure}

As an example, imagine a very massive pumped particle moving quickly
in a direction orthogonal to a cavity. The particle is initially
prepared as a very well localised wave packet. The pump is weak, so
that the atom does not feel any potential, but the coupling to the
cavity mode is strong, although not strong enough to create a big
field that would act back on the atom. In this case the cavity field
is weak, but is very sensitive to the position of the atom, on the
other hand, the atom does not feel the field at all. If the particle
is quick enough, it can travel several pump wavelengths before its
wave packet spreads noticeably. The cavity decay rate \(\kappa\) is
big enough so that the field follows adiabatically even this quick
atomic motion. In this case in the initial phase of the dynamics the
cavity field is almost a classical field scattered by an almost
classical point-like particle. This field we can calculate explicitly:
\begin{equation}
\label{eq:ClassicalField}
\avr{a}=\frac{\text{sign}(U_0)\sqrt{U_0\,\eta_\text{eff}}}{\DC-U_0+i\kappa}\zeta(x),
\end{equation}
where \(\zeta\) is the pump mode function, and \(x\) is the position of
the atom. An example for such a test run is displayed in
Fig.~\ref{fig:TestRun}.

\section*{Acknowledgement}
A.~V.~acknowledges support from the National Scientific Fund of
Hungary (Contract Nos.~T043079, T049234, NF68736).

\appendix

\section{Description of the MCWF method}
\label{sec:method}
The MCWF method \cite{carmichael87,dalibard92,dum92,molmer93} aims at the
simulation of open quantum systems based on a stochastic (``Monte
Carlo'') state vector. In terms of dimensionality, this is certainly
a huge advantage as compared to solving the Master equation
directly. On the other hand, stochasticity requires us to run many
trajectories, but the method provides an optimal sampling of the
ensemble density operator so that the relative error is inversely
proportional to the number of trajectories.

The optimal sampling is achieved by evolving the state vector in two
steps, one deterministic and one stochastic (quantum jump). Suppose
that the Master equation of the system is of the form
\begin{equation}
\label{eq:mm}
\dot\rho=\frac i\hbar\comm\rho H+
\Liou\rho\equiv\frac i\hbar\comm\rho H+
\sum_m\lp J_m\rho J_m^\dag-\frac12\comm{J_m^\dag J_m}{\rho}_+\rp,
\end{equation}
the usual form in quantum optics. At time \(t\) the system is in a
state with normalised state vector \(\ket{\Psi(t)}\). To obtain the
state vector at time \(t+\delta t\) up to first order in \(\delta t\):
\begin{enumerate}
\item The state vector is evolved according to the non-unitary
dynamics
\begin{equation}
\label{eq:DynnU}
i\hbar\frac{d\ket{\Psi}}{dt}=H_{\text{nH}}\ket{\Psi}
\end{equation}
with the non-Hermitian Hamiltonian
\begin{equation}
\label{eq:HamnH}
H_{\text{nH}}=H-\frac{i\hbar}2\sum_m J^\dag_m J_m
\end{equation}
to obtain (up to first order in \(\delta t\))
\begin{equation}
\label{eq:PsinH}
\ket{\Psi_{\text{nH}}(t+\delta t)}=
\lp1-\frac{iH_{\text{nH}}\,\delta t}\hbar\rp
\ket{\Psi(t)}.
\end{equation}
Since \(H_{\text{nH}}\) is non-Hermitian, this new state vector is not
normalised. The square of its norm reads
\begin{equation}
\braket{\Psi_{\text{nH}}(t+\delta t)}{\Psi_{\text{nH}}(t+\delta t)}\\=
\bra{\Psi(t)}\lp1+\frac{iH^\dag_{\text{nH}}\,\delta t}\hbar\rp
\lp1-\frac{iH_{\text{nH}}\,\delta t}\hbar\rp\ket{\Psi(t)}\equiv 1-\delta p,
\end{equation}
where \(\delta p\) reads
\begin{subequations}
\begin{align}
\delta p&=\delta t\,\frac i\hbar 
\bra{\Psi(t)} H_{\text{nH}}-H^\dag_{\text{nH}}\ket{\Psi(t)}\equiv\sum_m\delta p_m,\\
\delta p_m&=\delta t\,\bra{\Psi(t)} J^\dag_m J_m\ket{\Psi(t)}\geq 0.
\end{align}
\end{subequations}
Note that the time step \(\delta t\) should be small enough so that
this first-order calculation be valid. In particular, we require that 
\begin{equation}
\label{eq:dplimit}
\delta p\ll1.
\end{equation}

\item A possible quantum jump with total probability \(\delta p\). For
the physical interpretation of such a jump see eg Refs.\
\citep{dum92,molmer93}. We choose a random number \(\epsilon\) between
0 and 1, and if \(\delta p<\epsilon\), which should mostly be the
case, no jump occurs and for the new normalised state vector at
\(t+\delta t\) we take
\begin{equation}
\label{eq:renorm_ordodt}
\ket{\Psi(t+\delta t)}=
\frac{\ket{\Psi_{\text{nH}}(t+\delta t)}}{\sqrt{1-\delta p}}.
\end{equation}
If \(\epsilon<\delta p\), on the other hand, a quantum jump occurs,
and the new normalised state vector is chosen from among the different
state vectors
\(J_m\ket{\Psi(t)}\) according to the probability distribution
\(\Pi_m=\delta p_m/\delta p\):
\begin{equation}
\ket{\Psi(t+\delta t)}=
\sqrt{\delta t}\frac{J_m\ket{\Psi(t)}}{\sqrt{\delta p_m}}.
\end{equation}
\end{enumerate}

Obviously, however, we can and must do much better than this. Indeed,
assume that for some time no quantum jump occurs, and we perform Step
1 several times consecutively. This would be equivalent to evolving
the Schr\"odinger equation with the most naive first order (Euler)
method, which is known to be unstable and hence fail in most cases of
interest. In our framework, we choose to use instead an adaptive
step-size ODE routine, usually the embedded Runge-Kutta Cash-Karp
algorithm \cite{numrec}. In this case the time step is intrinsically
bounded by a precision requirement in the ODE stepper, but also by the
condition (\ref{eq:dplimit}), which is taken care of by our MCWF
stepper. Since in the ODE we are now much better than \(O(\delta t)\),
the renormalisation of the state vector is performed exactly rather
than to \(O(\delta t)\) as in Eq.~(\ref{eq:renorm_ordodt}).

In many situations it pays to use some sort of interaction picture,
which means that instead of Eq.~(\ref{eq:DynnU}) we strive to solve 
\begin{equation}
\label{eq:InU}
i\hbar\frac{d\ket{\Psi_{\text{I}}}}{dt}=U^{-1}\lp
H_{\text{nH}}U-i\hbar\frac{dU}{dt}\rp\ket{\Psi_{\text I}},
\end{equation}
where \(\ket{\Psi_{\text I}}=U^{-1}\ket\Psi\). Note that \(U\) can be
non-unitary. The two pictures are accorded after each time step, ie
before the time step \(\ket{\Psi_{\text I}(t)}=\ket{\Psi(t)}\) and
after the time step the transformation \(\ket{\Psi(t+\delta
t)}=U(\delta t)\ket{\Psi_{\text I}(t+\delta t)}\) is
performed. This we do on one hand for convenience and for
compatibility with the case when no interaction picture is used, but
on the other hand also because \(U(t)\) is non-unitary and hence for
\(t\to\infty\) some of its elements will become very large, while
others very small, possibly resulting in numerical problems. It is in
fact advisable to avoid evaluating \(U(t)\) with very large \(t\) arguments.

\section{Interacting systems --- State vector slices}
\label{sec:slices}
The main objective of the development of the present framework was to
allow users to compose composite systems at will from elementary
systems and interactions already provided in the framework, and
perform simulations for these composite systems. We can think of
quantum optics: several atoms of different structure interacting with
light fields or cavity modes. A concrete example is given in
Sec.~\ref{sec:CQED}.

Let us consider what we expect from an element of such a composite
system. This element will be a class, containing all the necessary
parameters specific to the given elementary system, and featuring eg a
function which calculates the effect of the free elementary-system
Hamiltonian \(H_{\text{at}}\) on a state vector. The
Hamiltonian \(H\) for a composite system of \(N\) subsystems in terms
of this Hamiltonian reads
\begin{equation}
\label{eq:CompositeH}
H=H_0+\dots+H_{\text{at}}+\dots+H_N+H^{\text{interaction}},
\end{equation}
The action of the elementary Hamiltonian \(H_{\text{at}}\) on a state vector
\(\ket\Psi\) expanded in a basis specified by some quantum numbers
\(\left\{i_n\right\}_{n=0\dots N}\) can be written as
\begin{equation}
\label{eq:dummy}
\bra{\left\{i_n\right\}_{n=0\dots N}}H_{\text{at}}\ket\Psi=
\sum_{j_{\text{at}}}
\lp H_{\text{at}}^{\text{elem}}\rp_{i_{\text{at}},j_{\text{at}}}
\braket{i_0,\dots,j_{\text{at}},\dots,i_N}\Psi.
\end{equation}

Since at the time of developing the class of the given elementary
system we do not know in which environment it will be embedded, we
expect the very same piece of code to work independently of the
environment. On the other hand, it has to know something about the
environment because as we see in Eq.~(\ref{eq:dummy}) the
multiplication by the matrix of \(H_{\text{at}}^{\text{elem}}\) has to
be performed for all possible combinations of the ``dummy'' quantum
numbers \(\left\{i_n\right\}_{n\neq\text{at}}\).

The state vector is ultimately stored as a one dimensional array (a
CPA) no matter how complex the system is, and the quantum numbers
\(\left\{i_n\right\}_{n=0\dots N}\) are mapped to a one dimensional
index by the indexing function
\begin{equation}
\label{eq:indexing}
I(i_0,\dots,i_N)=\sum_{n=0}^{N}i_n\prod_{n+1}^N d_m,
\end{equation}
where \(d\) denotes the dimension of the subsystem. Hence, the
information needed by \(H_{\text{at}}\) about the environment can be
condensed into the concept of \emph{array slices}, which, in our
framework is implemented by the \texttt{CPA\_View} class. For a free
system, a \texttt{CPA\_View} class consists of an array
\texttt{firstS} which contains the indices
\(I(i_0,\dots,i_{\text{at}}=0,\dots,i_N)\) for all the possible
combinations of the dummies \(\left\{i_n\right\}_{n\neq\text{at}}\)
and an integer \(\texttt{stride}=\prod_{\text{at}+1}^N d_m\).

To each element of the array \texttt{firstS} of a \texttt{CPA\_View}
there corresponds a \texttt{CPA\_Slice} which contains one single index
\texttt{first} and the integer \texttt{stride}. One can say that
\texttt{CPA\_Slice} is the iterator type of \texttt{CPA\_View}. The
index corresponding to a subsystem quantum number \(i_{\text{at}}\) for a
given set of the dummy quantum numbers can then be calculated from the
slice alone as
\begin{equation}
I\lp i_{\text{at}} | \left\{i_n\right\}_{n\neq\text{at}}\rp
=\texttt{first}+\texttt{stride}\times i_{\text{at}}.
\end{equation}

All the environment-independent implementation of \(H_{\text{at}}\) and
eventually that of every operator acting on a subsystem \(at\) of a
composite system has to see from the environment is a
\texttt{CPA\_View}. Having received a \texttt{CPA\_View} as a
parameter all an elementary Hamiltonian \(H_{\text{at}}\) has to do is to
iterate over the dummy indices condensed into \texttt{firstS} and
apply the same matrix \(H_{\text{at}}^{\text{elem}}\) on the state-vector
slice specified by the corresponding \texttt{CPA\_Slice}. This concept
is realized by \texttt{H} and \texttt{H\_elem}, cf Sec.~\ref{sec:elem}.

\texttt{CPA\_View} is essentially an array of \texttt{CPA\_Slice}s, we
just save resources by storing \texttt{stride}, which is the
characteristic of the given subsystem embedded in the given
environment, only once.

As discussed in Sec.~\ref{sec:elem}, interactions are also
``elements'' in our framework. An interaction Hamiltonian operates on
several subsystems, therefore its \texttt{CPA\_View} has to contain as
many \texttt{stride}s, each corresponding to the stride characteristic
for the given subsystem in the given embedding environment.

\begin{figure}
\centering
\includegraphics[width=15cm]{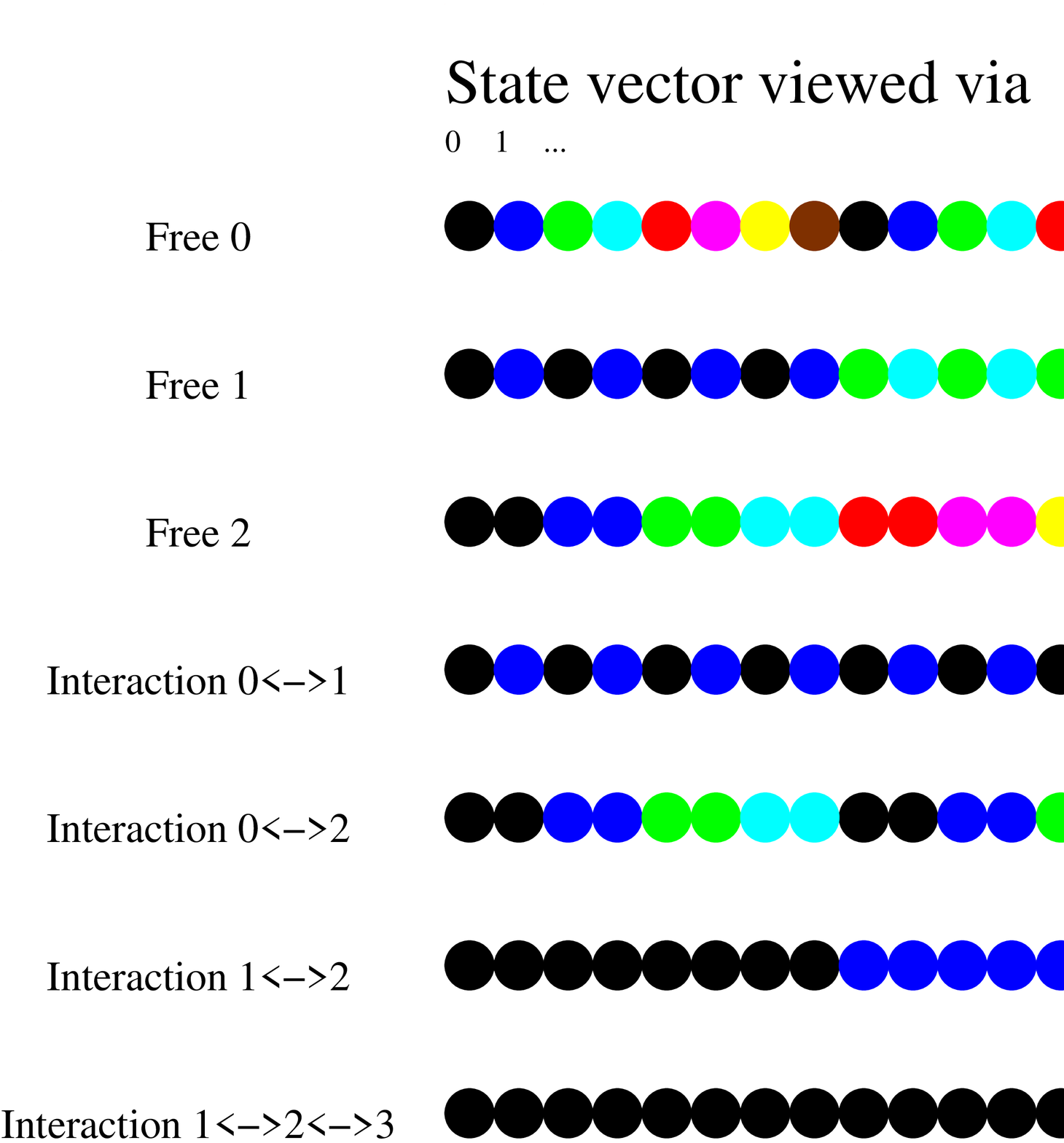}
\caption{The state vector of a system consisting of three subsystems
  with dimensions 3, 4 and 2, covered by different sets of
  \texttt{CPA\_Slice}s corresponding to the free subsystems and the
  interactions between the subsystems. One slice is the set
  of indices displayed in the same colour. A \texttt{CPA\_View} is
  essentially an array of slices with the modification that the
  \texttt{strideS} are stored only once.}
\label{fig:slice}
\end{figure}

The concept of array slices, the relationship between
\texttt{CPA\_Slice} and \texttt{CPA\_View}, and the fact that a
\texttt{CPA\_View} represents a way of looking on the state vector is
further exposed in Fig.~\ref{fig:slice}.

\begin{figure}
\centering
\listinginput{0}{example_driver1.C}
\caption{Full driver for one particle in a ring cavity sustaining two
  travelling-wave modes with opposite wave vectors, the \(-K\) mode
  being pumped. The definition of parameters (\texttt{omrec},
  \texttt{fin}, etc.) has been omitted for the sake of compactness.}
\label{fig:1pRingCav}
\end{figure}

\begin{figure}
\centering
\listinginput{0}{example_driver2.C}
\caption{The essential part of the driver for two identical pumped
  particles moving orthogonal to the axis of a cavity sustaining one
  single sinusoidal mode --- or otherwise, two identical particles
  moving in a one dimensional optical lattice with the cavity aligned
  orthogonally to the lattice.}
\label{fig:2p1m}
\end{figure}

\begin{figure}
\centering
\includegraphics[width=17cm]{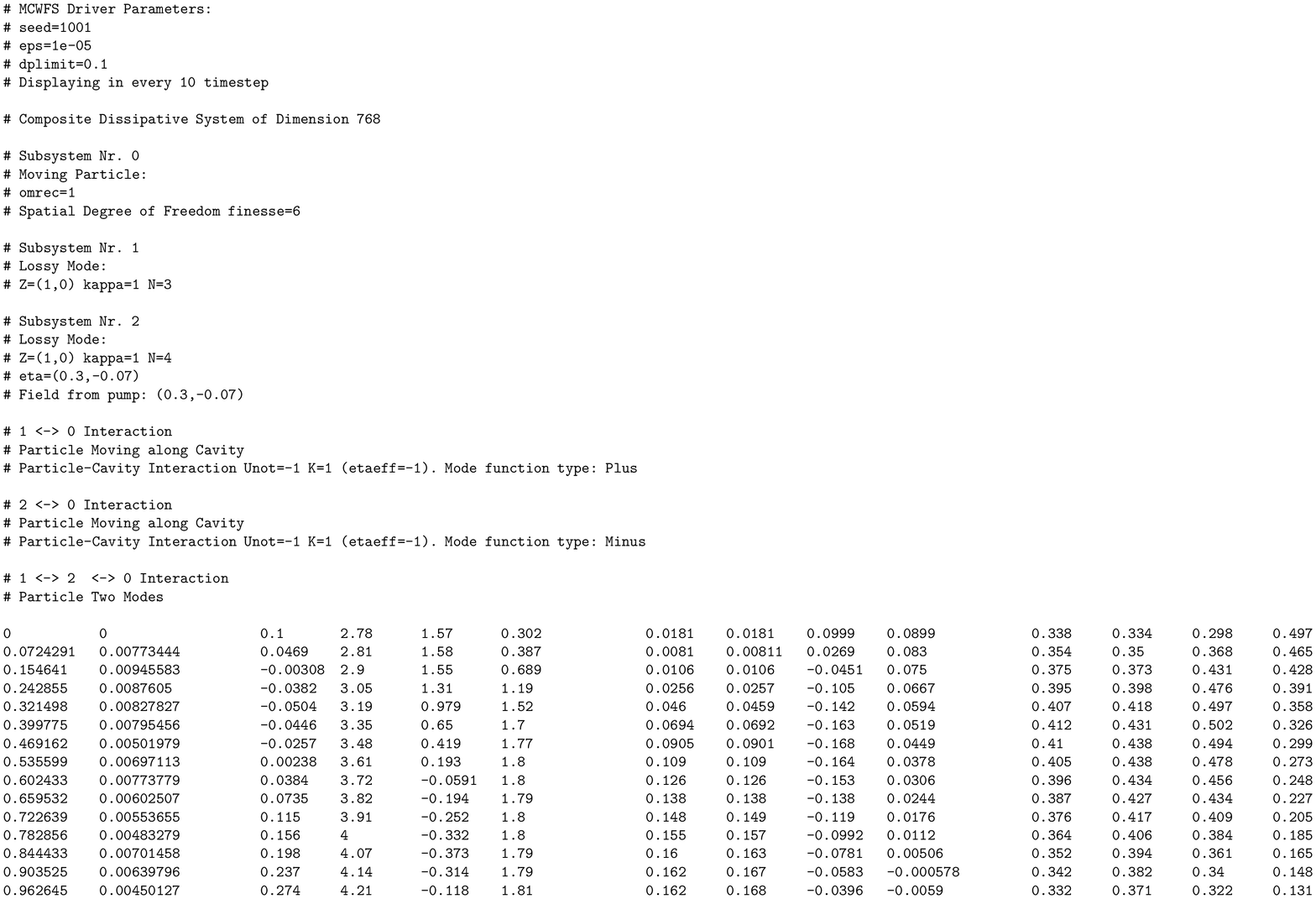}
\caption{Typical output of the ring-cavity driver of
  Fig.~\ref{fig:1pRingCav}. The first two columns are time and time
  step, respectively, then, separated by tab characters, the data
  stemming from the different subsystems follows: columns 3-6 contain
  the data from subsystem Nr.~0 \texttt{MovingParticle}, columns 7-10
  and 11-14 from the two cavity modes. The interaction elements make
  no output in this example.}
\label{fig:exampleOutput}
\end{figure}

\begin{thebibliography}{18}
\expandafter\ifx\csname natexlab\endcsname\relax\def\natexlab#1{#1}\fi
\expandafter\ifx\csname bibnamefont\endcsname\relax
  \def\bibnamefont#1{#1}\fi
\expandafter\ifx\csname bibfnamefont\endcsname\relax
  \def\bibfnamefont#1{#1}\fi
\expandafter\ifx\csname citenamefont\endcsname\relax
  \def\citenamefont#1{#1}\fi
\expandafter\ifx\csname url\endcsname\relax
  \def\url#1{\texttt{#1}}\fi
\expandafter\ifx\csname urlprefix\endcsname\relax\def\urlprefix{URL }\fi
\providecommand{\bibinfo}[2]{#2}
\providecommand{\eprint}[2][]{\url{#2}}

\bibitem[{\citenamefont{Vukics et~al.}(2005)\citenamefont{Vukics, Janszky, and
  Domokos}}]{vukics05a}
\bibinfo{author}{\bibfnamefont{A.}~\bibnamefont{Vukics}},
  \bibinfo{author}{\bibfnamefont{J.}~\bibnamefont{Janszky}}, \bibnamefont{and}
  \bibinfo{author}{\bibfnamefont{P.}~\bibnamefont{Domokos}},
  \bibinfo{journal}{J. Phys. B: At. Mol. Opt. Phys.}
  \textbf{\bibinfo{volume}{38}}, \bibinfo{pages}{1453} (\bibinfo{year}{2005}).

\bibitem[{\citenamefont{Vukics and Domokos}(2005)}]{vukics05b}
\bibinfo{author}{\bibfnamefont{A.}~\bibnamefont{Vukics}} \bibnamefont{and}
  \bibinfo{author}{\bibfnamefont{P.}~\bibnamefont{Domokos}},
  \bibinfo{journal}{Phys. Rev. A} \textbf{\bibinfo{volume}{72}},
  \bibinfo{pages}{031401} (\bibinfo{year}{2005}).

\bibitem[{\citenamefont{Maschler et~al.}(2005)\citenamefont{Maschler, Ritsch,
  Vukics, and Domokos}}]{maschler05b}
\bibinfo{author}{\bibfnamefont{C.}~\bibnamefont{Maschler}},
  \bibinfo{author}{\bibfnamefont{H.}~\bibnamefont{Ritsch}},
  \bibinfo{author}{\bibfnamefont{A.}~\bibnamefont{Vukics}}, \bibnamefont{and}
  \bibinfo{author}{\bibfnamefont{P.}~\bibnamefont{Domokos}}
  (\bibinfo{year}{2005}), \bibinfo{note}{quant-ph/0512101}.

\bibitem[{\citenamefont{Vukics}(2006)}]{vukicsthesis}
\bibinfo{author}{\bibfnamefont{A.}~\bibnamefont{Vukics}}, Ph.D. thesis,
  \bibinfo{school}{University of Szeged, Szeged} (\bibinfo{year}{2006}),
  \urlprefix\url{http://optics.szfki.kfki.hu/~vukics/thesispdf.pdf}.

\bibitem[{\citenamefont{Domokos and Ritsch}(2003)}]{domokos03}
\bibinfo{author}{\bibfnamefont{P.}~\bibnamefont{Domokos}} \bibnamefont{and}
  \bibinfo{author}{\bibfnamefont{H.}~\bibnamefont{Ritsch}},
  \bibinfo{journal}{J. Opt. Soc. Am. B} \textbf{\bibinfo{volume}{20}},
  \bibinfo{pages}{1098} (\bibinfo{year}{2003}).

\bibitem[{\citenamefont{Molmer et~al.}(1993)\citenamefont{Molmer, Castin, and
  Dalibard}}]{molmer93}
\bibinfo{author}{\bibfnamefont{K.}~\bibnamefont{Molmer}},
  \bibinfo{author}{\bibfnamefont{Y.}~\bibnamefont{Castin}}, \bibnamefont{and}
  \bibinfo{author}{\bibfnamefont{J.}~\bibnamefont{Dalibard}},
  \bibinfo{journal}{J. Opt. Soc. Am. B} \textbf{\bibinfo{volume}{10}},
  \bibinfo{pages}{524} (\bibinfo{year}{1993}).

\bibitem[{\citenamefont{Schack and Brun}(1996)}]{schack96}
\bibinfo{author}{\bibfnamefont{R.}~\bibnamefont{Schack}} \bibnamefont{and}
  \bibinfo{author}{\bibfnamefont{T.~A.} \bibnamefont{Brun}}
  (\bibinfo{year}{1996}), \bibinfo{note}{quant-ph/9608004}.

\bibitem[{\citenamefont{Tan}(1999)}]{tan99}
\bibinfo{author}{\bibfnamefont{S.~M.} \bibnamefont{Tan}}, \bibinfo{journal}{J.
  Opt. B: Quant. Semiclass. Opt.} \textbf{\bibinfo{volume}{1}},
  \bibinfo{pages}{424} (\bibinfo{year}{1999}).

\bibitem[{\citenamefont{Collecutt and Drummond}(2001)}]{collecutt01}
\bibinfo{author}{\bibfnamefont{G.}~\bibnamefont{Collecutt}} \bibnamefont{and}
  \bibinfo{author}{\bibfnamefont{P.~D.} \bibnamefont{Drummond}},
  \bibinfo{journal}{Comp. Phys. Commun.} \textbf{\bibinfo{volume}{142}},
  \bibinfo{pages}{219} (\bibinfo{year}{2001}).

\bibitem[{\citenamefont{Galassi et~al.}(2006)\citenamefont{Galassi, Davies,
  Theiler, Gough, Jungman, Booth, and Rossi}}]{gsl}
\bibinfo{author}{\bibfnamefont{M.}~\bibnamefont{Galassi}},
  \bibinfo{author}{\bibfnamefont{J.}~\bibnamefont{Davies}},
  \bibinfo{author}{\bibfnamefont{J.}~\bibnamefont{Theiler}},
  \bibinfo{author}{\bibfnamefont{B.}~\bibnamefont{Gough}},
  \bibinfo{author}{\bibfnamefont{G.}~\bibnamefont{Jungman}},
  \bibinfo{author}{\bibfnamefont{M.}~\bibnamefont{Booth}}, \bibnamefont{and}
  \bibinfo{author}{\bibfnamefont{F.}~\bibnamefont{Rossi}},
  \emph{\bibinfo{title}{GNU Scientific Library --- Reference Manual}},
  \bibinfo{organization}{GNU} (\bibinfo{year}{2006}),
  \urlprefix\url{http://www.gnu.org/software/gsl/manual/}.

\bibitem[{\citenamefont{Stroustrup}(1997)}]{stroustrup}
\bibinfo{author}{\bibfnamefont{B.}~\bibnamefont{Stroustrup}},
  \emph{\bibinfo{title}{The C++ Programming Language}}
  (\bibinfo{publisher}{Addison-Wesley}, \bibinfo{year}{1997}),
  \bibinfo{edition}{3rd} ed.

\bibitem[{\citenamefont{Gisin and Percival}(1992)}]{gisin92}
\bibinfo{author}{\bibfnamefont{N.}~\bibnamefont{Gisin}} \bibnamefont{and}
  \bibinfo{author}{\bibfnamefont{I.~C.} \bibnamefont{Percival}},
  \bibinfo{journal}{J. Phys. A} \textbf{\bibinfo{volume}{25}},
  \bibinfo{pages}{5677} (\bibinfo{year}{1992}).

\bibitem[{\citenamefont{Diosi}(1986)}]{diosi86}
\bibinfo{author}{\bibfnamefont{L.}~\bibnamefont{Diosi}},
  \bibinfo{journal}{Phys. Lett. A} \textbf{\bibinfo{volume}{114}},
  \bibinfo{pages}{451} (\bibinfo{year}{1986}).

\bibitem[{\citenamefont{Vukics et~al.}(2004)\citenamefont{Vukics, Domokos, and
  Ritsch}}]{vukics04}
\bibinfo{author}{\bibfnamefont{A.}~\bibnamefont{Vukics}},
  \bibinfo{author}{\bibfnamefont{P.}~\bibnamefont{Domokos}}, \bibnamefont{and}
  \bibinfo{author}{\bibfnamefont{H.}~\bibnamefont{Ritsch}},
  \bibinfo{journal}{J. Opt. B: Quant. Semiclass. Opt.}
  \textbf{\bibinfo{volume}{6}}, \bibinfo{pages}{143} (\bibinfo{year}{2004}).

\bibitem[{\citenamefont{Carmichael}(1987)}]{carmichael87}
\bibinfo{author}{\bibfnamefont{H.~J.} \bibnamefont{Carmichael}},
  \bibinfo{journal}{J. Opt. Soc. Am. B} \textbf{\bibinfo{volume}{4}},
  \bibinfo{pages}{1588} (\bibinfo{year}{1987}).

\bibitem[{\citenamefont{Dalibard et~al.}(1992)\citenamefont{Dalibard, Castin,
  and Molmer}}]{dalibard92}
\bibinfo{author}{\bibfnamefont{J.}~\bibnamefont{Dalibard}},
  \bibinfo{author}{\bibfnamefont{Y.}~\bibnamefont{Castin}}, \bibnamefont{and}
  \bibinfo{author}{\bibfnamefont{K.}~\bibnamefont{Molmer}},
  \bibinfo{journal}{Phys. Rev. Lett.} \textbf{\bibinfo{volume}{68}},
  \bibinfo{pages}{580} (\bibinfo{year}{1992}).

\bibitem[{\citenamefont{Dum et~al.}(1992)\citenamefont{Dum, Zoller, and
  Ritsch}}]{dum92}
\bibinfo{author}{\bibfnamefont{R.}~\bibnamefont{Dum}},
  \bibinfo{author}{\bibfnamefont{P.}~\bibnamefont{Zoller}}, \bibnamefont{and}
  \bibinfo{author}{\bibfnamefont{H.}~\bibnamefont{Ritsch}},
  \bibinfo{journal}{Phys. Rev. A} \textbf{\bibinfo{volume}{45}},
  \bibinfo{pages}{4879} (\bibinfo{year}{1992}).

\bibitem[{\citenamefont{Press et~al.}(1992)\citenamefont{Press, Teukolsky,
  Vetterling, and Flannery}}]{numrec}
\bibinfo{author}{\bibfnamefont{W.~H.} \bibnamefont{Press}},
  \bibinfo{author}{\bibfnamefont{S.~A.} \bibnamefont{Teukolsky}},
  \bibinfo{author}{\bibfnamefont{W.~T.} \bibnamefont{Vetterling}},
  \bibnamefont{and} \bibinfo{author}{\bibfnamefont{B.~P.}
  \bibnamefont{Flannery}}, \emph{\bibinfo{title}{Numerical Recipes in C}}
  (\bibinfo{publisher}{Cambridge}, \bibinfo{year}{1992}),
  \urlprefix\url{http://www.nr.com/}.

\end{thebibliography}
\end{document}